\documentclass[aip,amsmath,amssymb,reprint,]{revtex4-1}

\usepackage{graphicx}% Include figure files
\usepackage{dcolumn}% Align table columns on decimal point
\usepackage{bm}% bold math
\usepackage[utf8]{inputenc}
\usepackage[T1]{fontenc}
\usepackage{mathptmx}

\usepackage{physics}
\usepackage{amssymb}   % for math
\usepackage{amsmath}
\usepackage{multirow}
\usepackage{float}
\usepackage{graphicx}
\usepackage{xcolor}
\usepackage{hyperref}
\begin{document}
	
	%%%% Article title to be placed here
	\title{Rayleigh-B\'{e}nard magnetoconvection with temperature modulation}
	
\author{S. Hazra}
\altaffiliation[Also at ]{Department of Physics, Indian Institute of Technology Kharagpur, Kharagpur-721302, West Bengal, India}%Lines break automatically or can be forced with \\
\author{K. Kumar}%
\email{kumar@phy.iitkgp.ernet.in}
\affiliation{ 
	Department of Physics, Indian Institute of Technology Kharagpur, Kharagpur-721302, West Bengal, India}%

\author{S. Mitra}
\affiliation{Universit\'e Paris-Saclay, CNRS, Laboratoire de Physique des Solides, 91405 Orsay, France}
	%%%%% Subject entries to be placed here %%%%
	%\subject{Magnetoconvection, Parametric waves, Floquet theory}
	
	%%%% Keyword entries to be placed here %%%%
	%\keywords{Parametric waves, Rayleigh-B\'{e}nard magnetoconvection, Temperature modulation, Bi-critical points}
	
	%%%% Insert corresponding author and its email address}
%	\corres{Krishna Kumar\\
%		\email{kumar.phy.iitkgp@gmail.com}}
	
	%%%% Abstract%%%%%%%%%%%%
\begin{abstract}
Floquet analysis of modulated magnetoconvection in Rayleigh-B\'{e}nard\ geometry is performed.  The temperature of the lower plate is varied sinusoidally in time about a finite mean. As the Rayleigh number $\mathrm{Ra}$ is made to cross a critical value $\mathrm{Ra}_o$, the oscillatory magnetoconvection begins. The flow at the onset of magnetoconvection may oscillate either subharmonically or harmonically with the external modulation. The critical Rayleigh number $\mathrm{Ra}_o$ varies non-monotonically with $\omega$ for appreciable value of $a$. The temperature modulation may either postpone or prepone the appearance of magnetoconvection. The magnetoconvective flow always oscillates harmonically at larger values of $\omega$. The threshold $\mathrm{Ra}_o$ and the corresponding wave number $k_o$ approach to their values for the stationary magnetoconvection in the absence of modulation ($a = 0$), as $\omega \rightarrow \infty$. Two different zones of harmonic instability merge to form a single instability zone with two local minima for higher values of Chandrasekhar's number $\mathrm{Q}$, which is qualitatively new. We have also observed a new type of bicritical point, which involves two different sets of harmonic oscillations. The effects of variation of $\mathrm{Q}$ and $\mathrm{Pr}$ on the threshold $\mathrm{Ra}_o$ and critical wave number $k_o$ are also investigated.
%%%%%%%%%%%%%%%%
%%% Insert text which can be accommodated on first page in the tag "fmtext" %%%
%%%%%%%%%%%%%%% End of first page %%%%%%%%%%%%%%%%%%%%%	
\end{abstract}	
\maketitle

\section{Introduction}
Parametrically driven waves were first investigated by Faraday~\cite{Faraday_1831}. He studied the generation of standing surface waves in a liquid 
resting on a horizontal plate vibrating vertically. These waves, also known as Faraday waves~\cite{Miles&Henderson_1990,Kumar&Tuckerman_1994}, oscillate subharmonically with respect to the imposed vibration. The parametrically forced surface waves, which oscillate harmonically (synchronously) with the external vibration, was also predicted in a thin sheet of a viscous liquid~\cite{Kumar_1996} and observed in experiments~\cite{Mueller_etal_1997}. A bicritical point said to occur, when the harmonic and subharmonic waves are simultaneously generated~\cite{Kumar_1996,Wagner_etal_2003}. Faraday experiments show interesting fluid patterns~\cite{Wagner_etal_2003,Douady_1990,Fauve_etal_1992,Edwards&Fauve_1994,Kumar&Bajaj_1995,Kudrolli_Gollub_1996,Ibrahim_2015,Maity_etal_2020}.
Generation of bulk waves due to oscillatory flow under modulation was first observed in a Couette-Taylor flow~\cite{Donnelly_etal_1962,Donnelly_1964}, when the rotation rate of the inner cylinder was subjected to a time-periodic modulation. Such flows were investigated in a Rayleigh-B\'{e}nard system~\cite{Chandrasekhar_1961} under time-periodic modulation of (i) the temperature difference between two plates~
\cite{Venezian_1969,Rosenblat&Tanaka_1971,Yih&Li_1972,Gollub&Benson_1978,Ahlers_etal_1984,Roppo_etal_1984,Niemela&Donnelly_1987,Meyer_etal_1992,Smorodin&Luecke_2009,Smorodin&Luecke_2010,Singh_etal_2015,Kaur_etal_2016},  (ii) the acceleration due to gravity~\cite{Gresho&Sani_1970,Volmar&Mueller1997,Rogers_etal_2000}, and (iii) the external magnetic field~\cite{Belyaev-Smorodin_2009}.The excitation of harmonic  subharmonic oscillations of fluid flow in a modulated magnetoconvection is not investigated in  fluids of low Prandtl number, although the role of a constant magnetic field on parametrically forced surface waves in a metallic liquid was studied~\cite{Paul&Kumar_2007}. 
	
Results of Floquet analysis of modulated magnetoconvection~\cite{Proctor&Weiss_1982,Basak-Kumar_2016} in the  Rayleigh-B\'{e}nard geometry are presented. A sinusoidally varying temperature is imposed on the lower plate. The instability of the stationary conduction state is investigated against the periodic perturbations in fluids of low Prandtl number ($\mathrm{Pr} \le 0.1$). The time-periodic modulation forces the oscillatory magneto-convective flow at the  instability onset if $\mathrm{Ra}$ is made to cross a critical value $\mathrm{Ra}_o$. The magneto-convective flow oscillates either subharmonically or harmonically  with respect to the driving. For relatively lower values of the modulation frequency $\omega$, the threshold $\mathrm{Ra}_o$ for excitation of oscillatory magnetoconvection  varies in non-monotonic fashion and the critical wave number $k_o$ shows jumps. The parametrically forced magnetoconvection at the onset always oscillates harmonically (synchronously) with the temperature modulation at sufficiently higher values of $\omega$. The threshold for synchronous magnetoconvection, $\mathrm{Ra}_o$, monotonically decreases and settles at a value equal to the threshold for stationary Rayleigh-B\'{e}nard magnetoconvection, $\mathrm{Ra}_c (\mathrm{Q})$, as $\omega \rightarrow \infty$. The variation of $\mathrm{Ra}_o (\mathrm{Q})$ with $\mathrm{Q}$ is also non-monotonic, if the amplitude $a$ of modulation is appreciable. There is a new possibility at higher values of $\mathrm{Q}$: Instability zones in the $\mathrm{Ra}$-$k$ plane for excitation of harmonically oscillating waves located on the two sides of an instability zone for excitation of subharmonic waves  may collapse together to form a single instability zone. This leads to a qualitatively new type of bi-critical point, where harmonically oscillating magnetoconvective flows of two different wave numbers may be excited simultaneously. Besides, bi-critical points with simultaneous excitation of subharmonically and harmonically oscillating magnetoconvective flows at the instability onset are also possible.
 	
\section{Hydromagnetic System}
	
An electrically conducting incompressible viscous fluid of thickness $d$, volume expansion coefficient $\alpha$ and magnetic permeability $\mu$ is enclosed between two horizontal plates.
The viscous, thermal and magnetic diffusion coefficients of the fluid  are $\nu$, $\kappa$ and $\eta$, respectively. Dimensionless parameters $\mathrm{Pr} = \frac{\nu}{\kappa}$ and  $\mathrm{Pm}=\frac{\nu}{\eta}$ are called the thermal and the magnetic Prandtl numbers of the fluid, respectively. The horizontal plates are located at $z = -\frac{d}{2}$ and $z = +\frac{d}{2}$ in the coordinate system chosen. The temperature of the lower plate, $T_l$ is varying sinusoidally with time, i.e., $T_l \equiv T(z = -\frac{d}{2}) = T_1 + \tilde{a} \cos{\tilde{\omega} \tilde{t}}$. The upper plate is maintained at a temperature $T_u \equiv T(z = +\frac{d}{2}) = T_2 < T_1$. This imposes an adverse temperature gradient $\beta = (T_1-T_2)/d = \Delta T/d$ between the two plates for $\tilde{a}=0$. A spatially uniform and temporally constant magnetic field $\bm{B}_0 = B_0 {\bf \hat{z}}$ is also applied across the fluid. Figure~{1} depicts the system schematically.
	
\begin{figure}[H]
 \includegraphics[height=!, width=8.0cm]{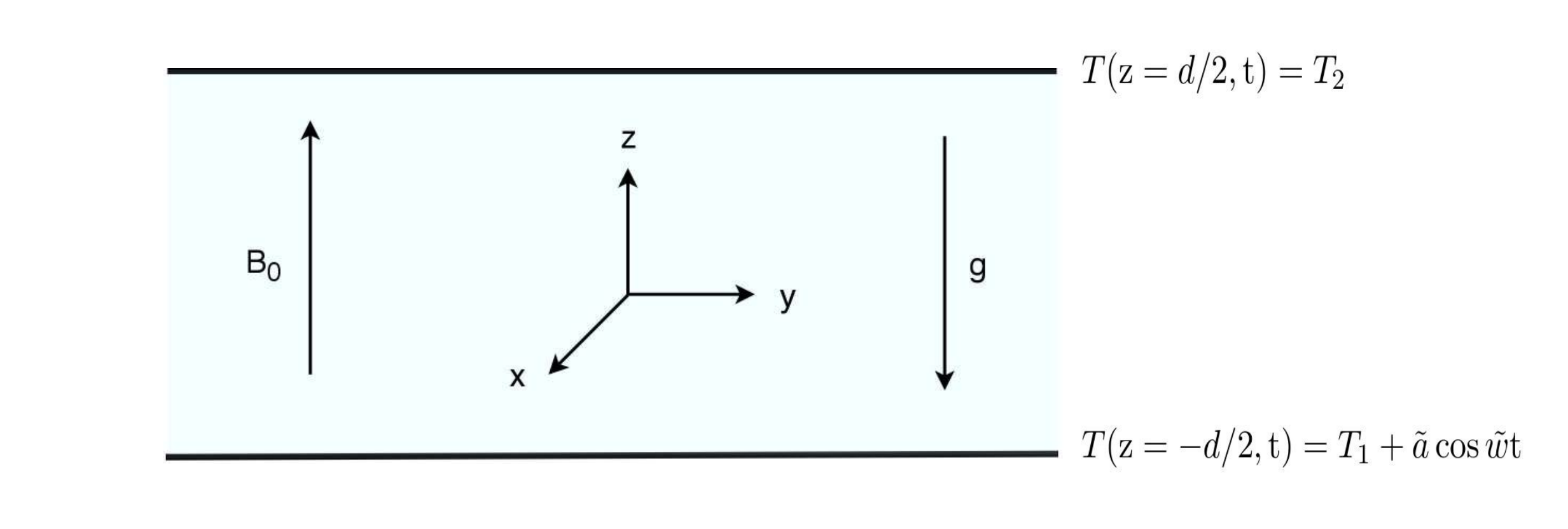} 
  \label{Schematic_diagram}
   \caption{(Colour online) Diagrammatic representation of the hydromagnetic system.}
	\end{figure}
	
The fluid is at rest in the basic sate. The fluid temperature $T_{cond}(z,t)$ in the basic state of stationary magnetoconduction is given as:
\begin{equation}
T_{cond} (z,t) = T_m -\beta z - \real{\Big\{\frac{\sinh{{\tilde{q}d\large(\frac{z}{d}-\frac{1}{2}\large)}}}{\sinh{\tilde{q}d}}  \tilde{a}e^{i\tilde{\omega} t}\Big\}}, \label{Tcond}
\end{equation}
where $T_m = (T_1+T_2)/2$ is the average fluid temperature in the basic state of magnetoconduction. The symbol $\real \{~\}$ stands the real part of its argument, and ${\tilde{q}}^2 = i\tilde{\omega}/\kappa$. The fluid density  $\rho_{cond} (z, t)$ and fluid pressure $P_{cond} (z, t)$ in the basic state are:
\begin{eqnarray}
\rho_{cond}(z, t) &=& \rho_m \Big[ 1 + \alpha \beta z + \real \Big\{ 
\frac{\sinh{\tilde{q}d(\frac{z}{d}-\frac{1}{2})}}{\sinh{qd}}\alpha \tilde{a} e^{i\tilde{\omega} t} \Big\} \Big],\label{Rho_cond}\\
P_{cond} (z,t) &=& P_0 +\rho_m gd \Big[ \frac{1}{2}-\frac{z}{d} +\frac{\alpha \beta d}{2} \Big(\frac{1}{4}-\frac{z^2}{d^2}\Big) \nonumber\\
&+&\real{ \Big\{ \Big( \frac{ 1-\cosh{\tilde{q}d(\frac{1}{2}-\frac{z}{d})}}{\sinh{\tilde{q}d}} \Big) \frac{\alpha\tilde{a}}{\tilde{q}d} e^{i\tilde{\omega} t}\Big\}} \Big]-\frac{B_0^2}{2\mu},\label{Pcond}
\end{eqnarray}
where $\rho_m$ is a reference density defined at the temperature $T_m$, and $P_0$ is a constant pressure at the upper plate. In the limit $\omega \rightarrow 0$, the temperature, density and the pressure fields in the magnetoconductive state are given as: 
\begin{eqnarray}
T_{cond} (z,\omega \rightarrow 0) &=& T_m -\beta z - \tilde{a}\large(\frac{z}{d}-\frac{1}{2}\large),\label{Tcondw0}\\ 
\rho_{cond}(z, \omega \rightarrow 0) &=& \rho_m \Big[ 1 + \alpha \beta z + \alpha \tilde{a} \large(\frac{z}{d}-\frac{1}{2}\large)\Big],\label{Rho_condw0}\\
P_{cond} (z,\omega \rightarrow 0) &=& P_0 +\rho_m gd\Big[\frac{1}{2}-\frac{z}{d} + \frac{\alpha \beta d}{2} \Big(\frac{1}{4}-\frac{z^2}{d^2}\Big) \nonumber \\
&-&\frac{\alpha \tilde{a}}{2} {\Big(\frac{1}{2}-\frac{z}{d} \Big)}^2 \Big)\Big]-\frac{B_0^2}{2\mu} \label{Pcondw0}
\end{eqnarray}
As the temperature gradient $\beta$ becomes greater than a threshold value for given set of values for $\tilde{a}$, $\tilde{\omega}$, $\mathrm{Pr}$ and $\boldsymbol{B}_0$, the magnetoconvection sets in. The velocity field becomes non-zero ($\tilde{\boldsymbol{u}} \neq 0$) and all other fields are perturbed.
\begin{eqnarray}
T_{cond} (z,t) \rightarrow T ({\bf r}, t) &=& T_{cond} (z,t) +\tilde{\theta} ({\bf r}, t),\\
\rho_{cond} (z,t) \rightarrow \rho({\bf r}, t) &=& \rho_{cond} (z,t) + \rho_m\alpha \tilde{\theta},\\
P_{cond} (z,t) \rightarrow P ({\bf r}, t) &=& P_{cond} (z,t) +\tilde{p} ({\bf r}, t),\\
\bm{B}_0 \rightarrow \bm{B} ({\bf r}, t) &=& \bm{B}_0 + \tilde{\bm{b}} ({\bf r}, t),
\end{eqnarray}
where ${\bf r} = x {\bf \hat{x}} + y {\bf \hat{y}}+ z {\bf \hat{z}}$. The fluid height $d$ is a natural length scale of the problem. The viscous diffusion time $d^2/\nu$ is chosen as the time scale. The choice of these two scales sets a scale for fluid velocity, which is the viscous diffusion speed $\nu/d$. Similarly, the quantities $\mathrm{Pr} \beta d$ and $\mathrm{Pm} B_0$ are chosen as typical scales for the temperature perturbation and the induced magnetic field, respectively. As the magnetic Prandtl number is of the order $10^{-5}$ or less for a earthly fluid, we set $\mathrm{Pm} = 0$ in the relevant equations of magnetoconvection. The linear version of the full dimensionless hydrodynamic equations describing the magnetoconvection, in the Oberbeck-Boussinesq approximation~\cite{Oberbeck_1879,Boussenisq_1903} are then given as:
\begin{eqnarray}
\mathrm{Pr} \partial_t \theta &=& \nabla\ ^2 \theta + u_3\Big[ 1 + \real{\Big\{\frac{\cosh{q({\frac{1}{2}-z})}}{\sinh{q}} q  a e^{i\omega t}\Big\}} \Big],\label{temperature}\\
\partial _t \bm{u} &=& -\bm{\nabla} (p + \mathrm{Q}b_3) + \mathrm{Q} \partial_z \bm{b}+\nabla ^2 \bm{u} + \mathrm{Ra} \theta \bm{z},\label{v_conv}\\ 
\nabla ^2 \bm{b} &=& - \partial _z \bm{u},\label{slaved_b}\\
\bm{\nabla \cdot u} &=& \bm{\nabla \cdot b} =0.\label{continuity}
\end{eqnarray}
In the above, $\mathrm{Ra} = \alpha g \beta d^4/(\nu \kappa)$ is known as the Rayleigh number and $\mathrm{Q} = B_0^2 d^2/(\mu \rho_m \nu \eta)$ is called the Chandrasekhar's number. They are two dimensionless parameters, which control the dynamics of magnetoconvection. The dimensionless amplitude and frequency of the time-periodic temperature modulation are defined as $a=\tilde{a}/(\beta d)$ and  $\omega = \tilde{\omega}d^2/\nu$, respectively. Besides, $q=\tilde{q}d$ is also a dimensionless parameter. Operating by curl ($\bm{\nabla~\times}$) twice on Eq.~\ref{v_conv}, using Eqs.~\ref{slaved_b}, \ref{continuity} and then projecting the resulting equation on the vertical axis, we arrive at
\begin{equation}
[\nabla^2(\partial _t - \nabla^2) + \mathrm{Q} \partial_{zz}]u_3 =  \mathrm{Ra} \nabla_H ^2 \theta.\label{velocity}
\end{equation}
	
The horizontal plates are considered to be rigid, thermally and electrically conducting. So the velocity field must vanish on the plates. In addition, the temperature fluctuations in the fluid must vanish at the horizontal plates. As the induced magnetic field cannot cross the electrically conducting horizontal plates, $ b_3 = 0$. The boundary conditions for the horizontal velocities may also be converted to an additional boundary condition on the vertical velocity using the equation of continuity~(\ref{continuity}). The boundary conditions are then summarised as:
\begin{equation}
\theta = u_3 = \partial_z u_3=0~~\mbox{at}~~z= \pm \frac{1}{2}\label{bcs}
\end{equation}
Eqs.~\ref{temperature} and~\ref{velocity} and the boundary conditions~\ref{bcs}
constitute the linear stability problem for the modulated Rayleigh-B\'{e}nard magnetoconvection in terms of the field variables $\theta$ and $u_3$.
	
\section{Floquet Analysis}
The growth of perturbations just above the onset of magnetoconvection is governed by the hydomagnetic system defined by Eqs.~\ref{temperature}, \ref{velocity} \& \ref{bcs}. As the temperature of the lower plate varies periodically in time, the Floquet technique~\cite{Kumar&Tuckerman_1994, Kumar_1996}, adopted for a set of partial differential equations, is an appropriate method to investigate the stability of the stationary state of magneto-conduction. However, more complex solutions~\cite{Smorodin&Luecke_2009,Smorodin&Luecke_2010},  which are often observed in a nonlinear system are beyond the scope of this work. As $\mathrm{Ra}$ is made to cross a threshold $\mathrm{Ra}_o (\mathrm{Q}, \mathrm{Pr},a,\omega)$, the oscillatory magnetoconvection sets in.  All the perturbative fields are expanded as:
\begin{equation}
\begin{bmatrix}
	u_3\\
	\theta
	\end{bmatrix}
	=\sum_{n=1}^{N} \sum_{l=-L}^{L}
	\begin{bmatrix}
	w_{nl} \psi_n(z)\\
	\theta_{nl} \sin{n\pi (z+\frac{1}{2})}
	\end{bmatrix} e^{i(\bf{k}\cdot\bf{x})}e^{[s+i(l+\gamma)\omega]t},\label{expansion}
\end{equation}
where $\psi_n(z)$ is an element of orthogonal set of Chandrasekhar's functions~\cite{Chandrasekhar_1961}, which are defined as: 
\begin{eqnarray}
\psi_n (z) &=& \frac{\cosh{(\lambda_n z)}}{\cosh{(\lambda_n/2)}}-\frac{\cos{(\lambda_n z)}}{\cos{(\lambda_n/2)}}~~~\mbox{for odd}~n,~~\mbox{and}\\
\psi_n (z) &=& \frac{\sinh{(\lambda_n z)}}{\sinh{(\lambda_n/2)}}-\frac{\sin{(\lambda_n z)}}{\sin{(\lambda_n/2)}}~~~\mbox{for even}~n,
\end{eqnarray}
where $\lambda_n$ is the $n^{th}$ root of the transcendental equation given by,
\begin{equation}
\tanh{(\lambda_n/2)}+(-1)^{(n+1)}\tan{(\lambda_n/2)}=0.
\end{equation}
For odd and even values of $n$,  even and odd Chandrasekhar's functions are chosen. As the Chandrasekhar's functions and their first derivatives vanish at $z=\pm 1/2$, the expansions for velocity fields on the vertical coordinate $z$ are consistent with the boundary conditions. In eq.~\ref{expansion}, ${\bf k}$ $=$ $k_x {\bf \hat{x}}$ $+$ $k_y {\bf \hat{y}}$ is the wave vector in the horizontal plane and ${\bf x}$ $=$ $x {\bf \hat{x}}$ $+$ $y {\bf \hat{y}}$. Here, 
$k=\sqrt{k_x^2+k_y^2}$ is the two-dimensional wave-number of perturbations. The complex number $s+i\gamma\omega$ is the Floquet exponent. Real part $s$ of the Floquet exponent decides the temporal growth rate of all perturbative fields. The condition of marginal (neutral) stability is determined by setting $s=0$. Here, the number $\gamma$ can be either equal to $1/2$ or $1$. The fluid flow with $\gamma = 1/2$ is called {\it subharmonic} magnetoconvection, while those with $\gamma = 0$ is known as {\it harmonic} magnetoconvection.  Here $1 \le n \le N$ and $-L \le l \le +L$, where the integers N and L are, in principle, infinitely large.  The reality condition of the perturbative fields require that
\begin{eqnarray}
w_{n,-l} &=& w^*_{n,+l},~~\mbox{and}~~~~\theta_{n,-l}~~~~~=~~\theta^*_{n,+l}~~\mbox{for}~~\gamma = 0,\\
w_{n,-(l+1)} &=& w^*_{n,+l},~~~\mbox{and}~~\theta_{n,-(l+1)} =~~\theta^*_{n,+l}~~\mbox{for}~~\gamma = 1/2.\nonumber\\
\end{eqnarray}
In numerics, the values of N and L are chosen such that the eigenvalues have an error less than a preassigned value, which is chosen to be less than $10^{-4}$. Insertion of above expansions for the perturbative fields in Eqs.~\ref{temperature} and~\ref{velocity} leads to the following difference equation:
\begin{equation}
\bm{G}(l) \bm{\xi}(l) + Ra \bm{H} \bm{\xi}(l) = a \bm{M}[\bm{\xi}(l-1)+\bm{\xi}(l+1)],\label{diff_eq}
\end{equation}
where for each value of the integer $l$, $\bm{\xi}(l)$ is a column vector with $2N$ elements. The square matrix $\bm{G}(l)$ is of the size $2N \times 2N$ with elements denoted by $G_{nn'}$. Similarly, $H_{nn'}$ and $M_{nn'}$ are the elements of square matrices $\bm{H}$ and $\bm{M}$, respectively. The size of both the matrices $\bm{H}$ and $\bm{M}$ is $2N \times 2N$. The column vector $\bm{\xi}(l)$ and square matrices $\bm{G}(l)$, $\bm{H}$ and 
$\bm{M}$ are given as:
\begin{eqnarray}
 \bm{\xi}(l) &=& [w_{1}(l), w_{2}(l), ..., w_{N}(l);\theta_{1}(l), \theta_{2}(l), ..., \theta_{N}(l)]^T,\\
 \bm{G}(l) &=& \begin{bmatrix}
 \bm{A}(l) & \bm{O}\\
 \bm{C} & \bm{B}(l)
 \end{bmatrix},~~ 
 \bm{H}=\begin{bmatrix}
 \bm{O} & \bm{D}\\
 \bm{O} & \bm{O}
 \end{bmatrix},~~
 \bm{M}=\begin{bmatrix}
 \bm{O} & \bm{O}\\
 \bm{E} & \bm{O}
 \end{bmatrix},\nonumber\\
\end{eqnarray}
where $\bm{A}(l)$, $\bm{B}(l)$, $\bm{C}$, $\bm{D}$ and $\bm{E}$ are square matrices of size $N \times N$ for each value of the integer $l$ with their components defined as  $A_{nn'}(l)$,  $B_{nn'}(l)$, $C_{nn'}$, $D_{nn'}$ and $E_{nn'}$, respectively. These matrices are given as:
\begin{eqnarray}
A_{nn'}(l) &=& \Big[-k^2\{s+i(l+\gamma)\omega\}-k^4\Big]\delta_{nn'} -\int_{- \frac{1}{2}}^{\frac{1}{2}} \psi_n(z) \partial_{zzzz}\psi_{n'}(z) dz\nonumber\\
&+& \Big[\mathrm{Q}+2k^2+s+i(l+\gamma)\omega\Big]\int_{-\frac{1}{2}}^{\frac{1}{2}} \psi_n(z) \partial_{zz}\psi_{n'}(z) dz,\\
B_{nn'}(l) &=& \Big[\mathrm{Pr} \{s+i(l+\gamma)\omega \}+k^2+n^2\pi^2\Big]
\delta_{nn'},\\
C_{nn'} &=& -2 \delta_{nn'}, \\
D_{nn'} &=& \frac{k^2}{2}\delta_{nn'},\\
E_{nn'} &=& \real{\Big[\frac{q \mathrm{Pr}}{\sinh{q}}\int_{-\frac{1}{2}}^{\frac{1}{2}}  \psi_n(z) \psi_{n'}(z)\cosh{\{q({1/2-z)}\}} dz \Big]},
\end{eqnarray}
where $\delta_{nn'}$ stands for Kronecker delta, and $\bm{O}$ represents a null matrix of size $N \times N$. As $l$ varies from $-L$ to $+L$, there are $2L+1$ equations like the one given in Eq.~\ref{diff_eq} for harmonic case ($\gamma =0$) and $2L$ equations  for subharmonic case ($\gamma =1/2$). All these difference equations may be put in the form of a generalised eigenvalue equation, which is given as:
\begin{equation}
\bm{U}\bm{X} = \frac{1}{\mathrm{Ra}} \bm{V}\bm{X}.\label{eigenvalue_gen}
\end{equation}
For harmonic solution ($\gamma =0$), the column vector $\bm{X}$ $=$ $[... \bm{\xi}(-2) \bm{\xi}(-1) \bm{\xi}(0) \bm{\xi}(1) \bm{\xi}(2) ...]^T$ consists of $2L+1$ smaller column vectors $\bm{\xi}(l)$. In case of subharmonic solution, $\bm{X}$ $=$ $[... \bm{\xi}(-3)\bm{\xi}(-2) \bm{\xi}(-1) \bm{\xi}(0) \bm{\xi}(1) \bm{\xi}(2) ...]^T$ consists of $2L$ smaller column vectors $\bm{\xi}(l)$. In both the cases, $\bm{\xi}(l)$ has $2N$ elements for each $l$. The matrix $\bm{U}$ is a block diagonal with the square matrix $\bm{H}$ repeated $2L+1$ times along the diagonal for $\gamma = 0$ and $2L$ times for $\gamma = 1/2$. The square matrix $\bm{V}$ is a banded matrix with only the first sub-diagonal, the diagonal, and the first super-diagonal blocks non-zero. The diagonal elements consist of matrices $-\bm{G}(l)$ for different values of $l$. The sub-diagonal and super-diagonal elements consist of $a\bm{M}$ repeated appropriate number of times. For harmonic case, the banded matrix $\bm{V}$ has odd number of block matrices along its diagonal and it is given as:
\begin{equation}
	\bm{V} =
	\begin{bmatrix}
	.. &... &... &... &... &... &..\\
	.. &-\bm{G}(-2) &a\bm{M} &\bm{\tilde{O}} &\bm{\tilde{O}} &\bm{\tilde{O}} &..\\
	.. &a\bm{M} &-\bm{G}(-1) &a\bm{M} &\bm{\tilde{O}} &\bm{\tilde{O}} &..\\
	.. &\bm{\tilde{O}} &a\bm{M} &-\bm{G}(0) &a\bm{M} &\bm{\tilde{O}} &..\\
	.. &\bm{\tilde{O}} &\bm{\tilde{O}} &a\bm{M} &-\bm{G}(1) &a\bm{M} &..\\
	.. &\bm{\tilde{O}} &\bm{\tilde{O}} &\bm{\tilde{O}} &a\bm{M}  &-\bm{G}(2) &..\\
	.. &... &... &... &... &... &..
	\end{bmatrix}. 
\end{equation} 
The banded matrix $\bm{V}$ has even number of block matrices along its diagonal in case of subharmonic response and it is is given by,
\begin{equation}
	\bm{V} =
	\begin{bmatrix}
	.. &... &... &... &... &...\\
	.. &-\bm{G}(-2) &a\bm{M} &\bm{\tilde{O}} &\bm{\tilde{O}}  &..\\
	.. &a\bm{M} &-\bm{G}(-1) &a\bm{M} &\bm{\tilde{O}}  &..\\
	.. &\bm{\tilde{O}} &a\bm{M} &-\bm{G}(0) &a\bm{M}  &..\\
	.. &\bm{\tilde{O}} &\bm{\tilde{O}} &a\bm{M} &-\bm{G}(1)  &..\\
	.. &... &... &... &... &..
	\end{bmatrix},
\end{equation}
where $\bm{\tilde{O}}$ is a null matrix of size $2N \times 2N$. Operating by 
$\bm{V}^{-1}$ from left on both sides of Eq.~\ref{eigenvalue_gen}, it may be put in the form of a standard eigenvalue problem, which is given as:
\begin{equation}
\Big(\bm{V}^{-1}\bm{U}\Big)\bm{X} = \Big(\frac{1}{\mathrm{Ra}}\Big)\bm{X}. \label{eigenvalue}
\end{equation}
The threshold for the oscillatory magnetoconvection, $\mathrm{Ra}_o$, and the corresponding wave number, $k_o$, are computed by setting $s=0$. The eigenvalues of matrix $\bm{V}^{-1}\bm{U}$ give the possible values of $\frac{1}{\mathrm{Ra}}$ with all other parameters kept fixed. The largest eigenvalue of $\bm{V}^{-1}\bm{U}$ yields the lowest value of $\mathrm{Ra}$ for a set of values for $a$, $\omega$, $\mathrm{Q}$, $\mathrm{Pr}$ and $\gamma$. The marginal stability curve $\mathrm{Ra}(k)$ is plotted  by varying $k$ in small steps. The minimum of this curve yields the threshold $\mathrm{Ra}_o (a, \omega, \gamma, \mathrm{Q}, \mathrm{Pr})$ and critical wave-number $k_o (a, \omega, \gamma, \mathrm{Q}, \mathrm{Pr})$ at the onset of  oscillatory 
magnetoconvection for preassigned values of $a$, $\omega$, $\gamma$, $\mathrm{Q}$ and $\mathrm{Pr}$. By varying any one of the parameters $\omega$, $\mathrm{Q}$ and $\mathrm{Pr}$ in small steps, while keeping the other two at fixed values, we compute dependence of $\mathrm{Ra}_o$ and $k_o$ on $\omega$,  $\mathrm{Q}$ and $\mathrm{Pr}$ separately for harmonic as well as subharmonic oscillations.
	
\section{Results and Discussions}
	  
Magnetoconvection at the primary instability is known to be stationary~\cite{Chandrasekhar_1961} for $a=0$,  if $\mathrm{Pr} > \mathrm{Pm}$. To validate the computational technique used here, we compare critical values $\mathrm{Ra}_c (\mathrm{Q})$ and $k_c (\mathrm{Q})$ computed for the stationary magnetoconvection with those obtained from Chandrasekhar's linear theory.  We have chosen $N=10$ and $L=40$ for the purpose. Table~\ref{Table_1} shows a comparison of values determined by two methods. The maximum error in the value of $\mathrm{Ra}_{c} (\mathrm{Q})$ is $0.5\%$, while that in determination of $k_{c} (\mathrm{Q})$ is less than $0.2\%$. This method with sufficient terms in the expansion of perturbative fields is therefore expected to yield very accurate results for the critical values.
\begin{table}[H]
 \centering
  \begin{center}
   \caption{Comparison of the critical values, $\mathrm{Ra}_c (\mathrm{Q})$ and $k_c (\mathrm{Q})$, computed by the method presented here  with those evaluated from  Chandrasekhar's linear theory of stationary magnetoconvection.}
	\label{Table_1}
  \end{center}
	\begin{small}
	\hspace*{-5pt}
	\scalebox{0.9}{
	\begin{tabular}{|c|c|c|c|c|c|c|}
	\hline
\multirow{2}{*}{$\mathrm{Q}$} &\multicolumn{2}{|c|}{Chandrasekhar's results} &\multicolumn{2}{|c|}{Our results for $N=10$} &\multirow{2}{*}{$\%$ error in $\mathrm{Ra}_{c}\mathrm{(Q)}$}\\
&\multicolumn{2}{|c|}{} &\multicolumn{2}{|c|}{} &\\\cline{2-5}
					&$k_{c}$ &$Ra_{c}(Q)$ &$k_{c}$ &$Ra_{c}(Q)$ &\\
					\hline
					$0$ &$3.13$ &$1707.8$ &$3.13$ &$1708$ &$0.01\%$ \\
					$10$ &$3.25$ &$1945.9$ &$3.25$ &$1946$ &$0.01\%$ \\
					$50$ &$3.68$ &$2802.1$ &$3.68$ &$2802$ &$0.01\%$ \\
					$100$ &$4.00$ &$3757.4$ &$4.00$ &$3758$ &$0.02\%$ \\
					$200$ &$4.45$ &$5488.6$ &$4.45$ &$5493$ &$0.08\%$ \\
					$500$ &$5.16$ &$10110.0$  &$5.16$ &$10133$ &$0.23\%$ \\
					$1000$ &$5.80$ &$17103.0$ &$5.81$ &$17189$ &$0.50\%$\\\hline
	\end{tabular}}
	 \end{small}
	\end{table}
\begin{figure}[H]
	\includegraphics[height=!, width=8.0cm]{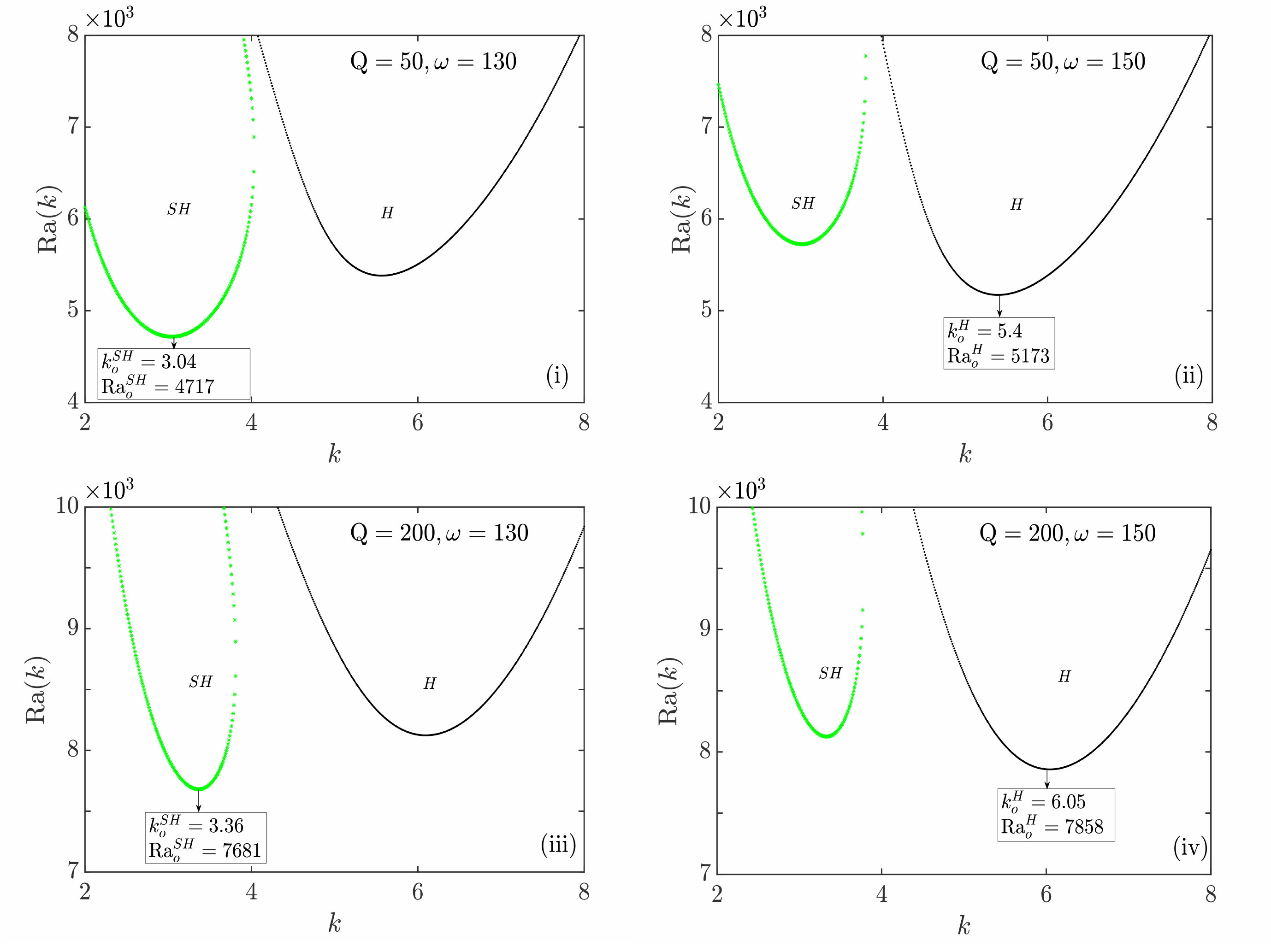}
	\caption{(Colour online) Instability zones for $Pr=0.1$ and $a=3$. Curves in the upper row are for $\mathrm{Q}=50$ with (i) $\omega=130$ and (ii) $\omega=150$. Curves in the lower row are for $\mathrm{Q} =200$ with (iii) $\omega=130$ and (iv) $\omega=150$. The region insider green (light grey) dots is the zone of subharmonic (SH) instability and the region inside black dots is the zone of harmonic (H) instability.}
	\label{Pr01_Rk}
\end{figure}

We now present the results of Floquet analysis for parametrically excited magnetoconvection. We have chosen $N=10$ and $L=40$. Figure~\ref{Pr01_Rk} shows the marginal stability curves for molten iron or Earth's liquid core ($\mathrm{Pr} = 0.1$). The dimensionless modulation amplitude $a$ is set to $3.0$. The upper left viewgraph shows two instability zones for $\mathrm{Q} = 50$ in the $\mathrm{Ra}$-$k$ plane: one for subharmonic (SH)  ($\gamma =1/2$) and another for harmonic (H) ($\gamma = 0$) magnetoconvection. They are the regions inside green (grey) and  black curves in the $\mathrm{Ra}$-$k$ plane, respectively. If a point ($\mathrm{Ra},k$) in this plane lies inside the subharmonic instability curve, the stationary conduction state becomes unstable to  subharmonically oscillating magnetoconvection under temperature modulation. Similarly, a point lying in side the harmonic  instability curve corresponds to the excitation of oscillatory magnetoconvection synchronous with the external modulation.  All points outside the marginal stability curves correspond to a state of stationary magnetoconduction. Tongue shaped curves, therefore, separate the regions of  conduction state, where the growth rate of perturbations is $s < 0$  from the regions of oscillatory instability, where  $s > 0$ in the $\mathrm{Ra}$-$k$ space. These curves are also known as neutral (marginal) stability curves. Perturbative fields neither grow nor decay for a set of values ($\mathrm{Ra}(k), k$) on these curves, as $s =0$ on them. Fig.~\ref{Pr01_Rk} (i) shows that the minimum of marginal curve for subharmonically excited waves is lower than that for harmonic waves for modulation frequency $\omega = 130$. Therefore, the parametrically excited flow for $\omega = 130$ oscillates subharmonically with the forcing at the primary instability. The threshold value of $\mathrm{Ra}$ and the corresponding $k$ value for $\omega=130$ are $\mathrm{Ra}_o^{SH} = 4717$ and $k_o^{SH} = 3.04$, respectively. As frequency is raised keeping $a$, $\mathrm{Q}$ and $\mathrm{Pr}$ fixed, the instability zone for $\gamma=1/2$ moves upward in the $\mathrm{Ra}$-$k$ plane, while the instability zone for $\gamma =0$ moves downward. The upward and downward shifts of the different instability zones are different. Fig.~\ref{Pr01_Rk} (ii) shows the first two instability zones for $\omega = 150$. In this case, the fluid flow excited at the onset of magnetoconvection is harmonic. The threshold for generating harmonically oscillating magnetoconvection at $\omega = 150$ is $\mathrm{Ra}_o^H = 5173$, and the excited wave number is $k_o^{H} = 5.40$. 
		
The lower row of Fig.~\ref{Pr01_Rk} displays stability zones for $\mathrm{Q} =200$, $\mathrm{Pr}=0.1$ and $a=3.0$. For modulation frequency $\omega = 130$, the magneto-convective flow at the primary instability oscillates subharmonically with respect to the imposed modulation [see Fig.~\ref{Pr01_Rk} (iii)]. The critical values of $\mathrm{Ra}$ and $k$ are:  $\mathrm{Ra}_o^{SH} = 7681$ and $k_o^{SH} = 3.36$ respectively. As the frequency is raised slightly above to $\omega = 150$, harmonically oscillating flow appears at the primary instability instead of subharmonic flow [see Fig.~\ref{Pr01_Rk} (iv)] with $\mathrm{Ra}_o^H = 7858$ and $k_o^{H} = 6.05$. The jump in wave-number is significant as subharmonic magneto-convective flow becomes harmonic. When the minima of two marginal stability curves are found to occur for the same value $\mathrm{Ra}$ in the $\mathrm{Ra}$-$k$ plane, a bi-critical point appears as soon as $\mathrm{Ra}$ is raised above $\mathrm{Ra}_o$. The fluid motion at the resulting bi-critical point involves subharmonic as well as harmonic oscillations. This kind of bi-critical point was predicted in the Faraday experiment  with a thin layer of viscous fluids~\cite{Kumar_1996}. They were also observed in experiments~\cite{Mueller_etal_1997,Wagner_etal_2003}.

The temperature, density and pressure become time independent if either $a \rightarrow 0$ or $\omega \rightarrow 0$. The corresponding critical values should correspond to those for the stationary magnetoconvection. For $a \rightarrow 0$, $\mathrm{Ra}_o (\mathrm{Q}, \mathrm{Pr}, a\rightarrow 0, \omega)$ $\rightarrow$ $\mathrm{Ra}_c (\mathrm{Q})$. However, due to  our choice of a cosine function for the temperature modulation of the lower plate, the imposed temperature difference, in the limit $\omega \rightarrow 0$, is ${\Delta T}_m (\omega \rightarrow 0) = \Delta T (1 + a)$ instead of $\Delta T$ (for $a=0$). The Rayleigh number now depends on ${\Delta T}_m$, which is larger than ${\Delta T}$ by a factor $(1+ {a})$ in the limit $\omega \rightarrow 0$.  Therefore, $\mathrm{Ra}_o (\mathrm{Q}, \mathrm{Pr}, a, \omega \rightarrow 0)$ approaches a fixed value $\mathrm{Ra}_s (\mathrm{Q})$, which would be less than the critical Rayleigh number for the stationary magnetoconvection $\mathrm{Ra}_c (\mathrm{Q})$ by a factor $(1+a)$. That is, $\mathrm{Ra}_o (\mathrm{Q}, \mathrm{Pr}, a, \omega\rightarrow 0)$ $\rightarrow$ $\mathrm{Ra}_s (\mathrm{Q})$ $= \mathrm{Ra}_{c}(\mathrm{Q})/(1+a)$. If the modulation amplitude $a$ and frequency $\omega$ both approach to zero simultaneously, then $\mathrm{Ra}_o (\mathrm{Q}, \mathrm{Pr}, a \rightarrow 0, \omega \rightarrow 0)$ approaches to  $\mathrm{Ra}_c (\mathrm{Q})$. For non-zero but small values of $\omega$, the value of $\mathrm{Ra}_o (\mathrm{Q}, \mathrm{Pr}, a, \omega)$ is to be determined numerically using the Floquet method. As shown below, $\mathrm{Ra}_o (\mathrm{Q}, \mathrm{Pr}, a, \omega)$ is found to vary non-monotonically for smaller values of $\omega$. Effects of modulation are confined a thin boundary layers near the plates at higher modulation frequencies. The thickness of these layers vanish as $\omega \rightarrow \infty$. Consequently, we expect $\mathrm{Ra}_o (\mathrm{Q}, \mathrm{Pr}, a, \omega\rightarrow \infty)$ $\rightarrow$ $\mathrm{Ra}_c (\mathrm{Q})$ for all values of $\mathrm{Pr}$ and $a$.

\begin{figure}[!]
	\includegraphics[height=!, width=8.0cm]{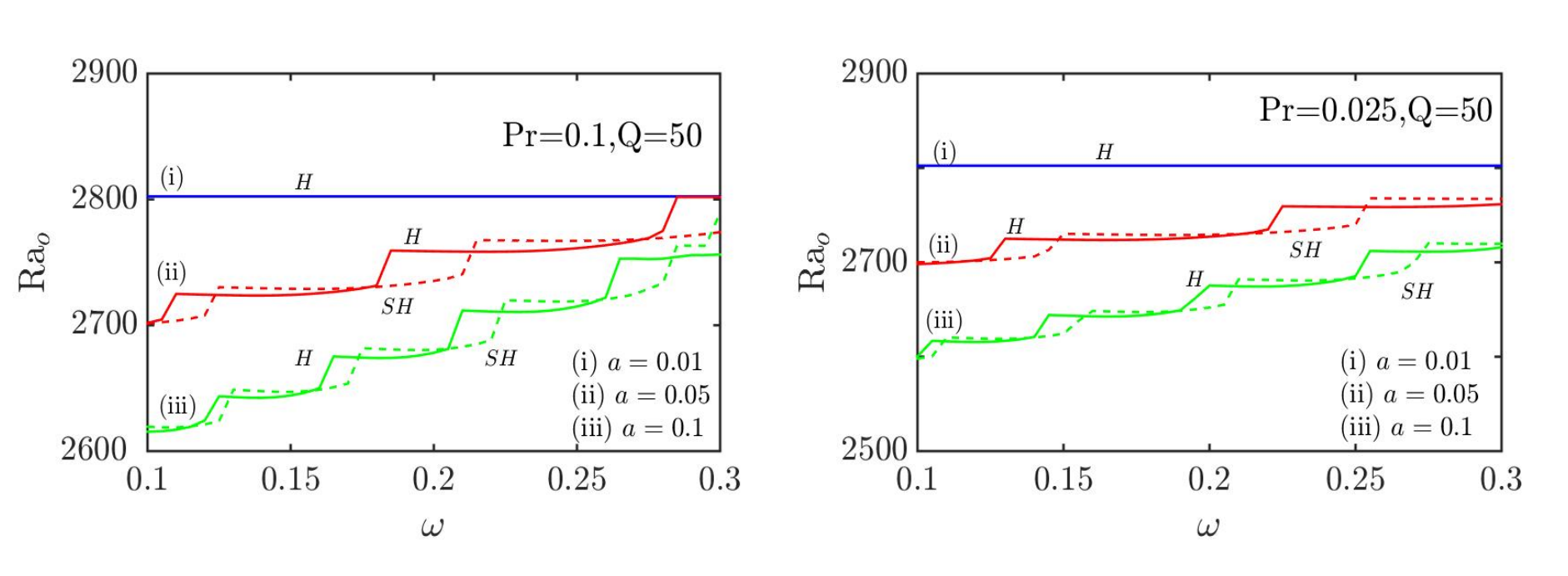}
	\caption{(Colour online) Variation of $\mathrm{Ra}_o (\omega)$ with the external frequency $\omega$ at very low values of $\omega$ for  $\mathrm{Q}=50$. The left and right view graphs are for $\mathrm{Pr}=0.1$ and $\mathrm{Pr}=0.025$, respectively. The magneto-convective motion oscillates either harmonically (solid curves) or subharmonically (dashed curves). Blue (black), red (grey) and light green (light grey) curves are for the dimensionless modulation amplitude $a$ equal to (i) $0.01$, (ii) $0.05$ and (iii) $0.1$, respectively.}
	\label{low_a_Q50}
\end{figure}

Figure~\ref{low_a_Q50} displays the variation of the critical Rayleigh number $\mathrm{Ra}_o (\omega)$  at very smaller values of $\omega$ for $\mathrm{Q}=50$. The left viewgraph is for Earth's liquid core ($\mathrm{Pr}=0.1$) and the right one for mercury ($\mathrm{Pr}=0.025$). The solid and dashed curves are for harmonic ($\gamma =0$) and subharmonic ($\gamma =1/2$) oscillations of the fluid flow. The value of $\mathrm{Ra}_o (\mathrm{Q}=50, a=0.01, \omega =0.1)$ is approximately $2800$ for both the cases (see the blue (black) curves). The oscillatory instability is harmonic in the frequency window $0.1 \le \omega \le 0.3$ for $a=0.01$. However, for $a=0.05$ (see the red (grey) curves) and $a=0.1$ (green (light grey) curves), the fluid motion at the instability onset shows alternately harmonic and subharmonic oscillations. The critical Rayleigh number, $\mathrm{Ra}_o$, for $\omega =0.1$ decreases as $a$ is increased for both the values of $\mathrm{Pr}$. That is, $\mathrm{Ra}_o (\mathrm{Q}=50, a=0.1, \omega =0.1)$ $<$ $\mathrm{Ra}_o (\mathrm{Q}=50, a=0.05, \omega =0.1)$ $<$ $\mathrm{Ra}_o (\mathrm{Q}=50, a=0.01,  \omega =0.1)$. The value of $\mathrm{Ra}_o (\mathrm{Q}, \mathrm{Pr}, a, \omega \rightarrow 0)$ approaches to a value $\mathrm{Ra}_s (\mathrm{Q}) = \mathrm{Ra}_c (\mathrm{Q})/ (1+a)$, as $\omega \rightarrow 0$. For very small values of $\omega$ ($\ll 1$), the elements $E_{n,n^{\prime}}$ of the square matrix $E$ take $0/0$ form as $\omega \rightarrow 0$. This makes the elements of the matrix $\bm{V}$ along the first sub-diagonal and super-diagonal of nearly indeterminate ($0/0$) form, which lead to larger numerical errors. The errors in computation of $\mathrm{Ra}_o$ at $\omega =0.1$ are within $1\%$ for $a \le 0.05$, while the same for $a=0.1$ is slightly less than $3\%$. For relatively larger values of $a$ ($ > 0.1$) and $\mathrm{Q}$, the eigenvalues of the matrix $\bm{V}$ may be computed quite accurately for dimensionless angular frequency $\omega > 1$.  However, the method used here works well for the actual modulation frequency $\tilde{f} \ge 10^{-2}$ Hz if the fluid thickness $d \ge 2$ mm. For a thicker layer of a fluid of smaller Prandtl number fluids, the threshold values may be computed accurately for much smaller values of $\tilde{f}$. We have presented here the data for liquids with $\mathrm{Pr} = 0.1$ and $0.025$. For a layer of liquid of $\mathrm{Pr} = 0.1$ (kinematic viscosity, $\nu = 6 \times 10^{-7}$ m$^2$/s at $2000$ K) of thickness $d$ varying from $2$ mm to $1$ cm at actual driving frequency of $1$ Hz, the dimensionless angular frequency $\omega$ varies approximately from  $40$ to $1050$. Similarly in a layer of mercury ($\mathrm{Pr} = 0.025$, $\nu = 1.14 \times 10^{-7}$ m$^2$/s at $303$ K) with  $2$ mm $\le d \le$ $1$ cm at the driving frequency $1$ Hz, $\omega$ varies approximately from $220$ to $5500$ approximately.

\begin{figure}[!]
	\includegraphics[height=!, width=8.0cm]{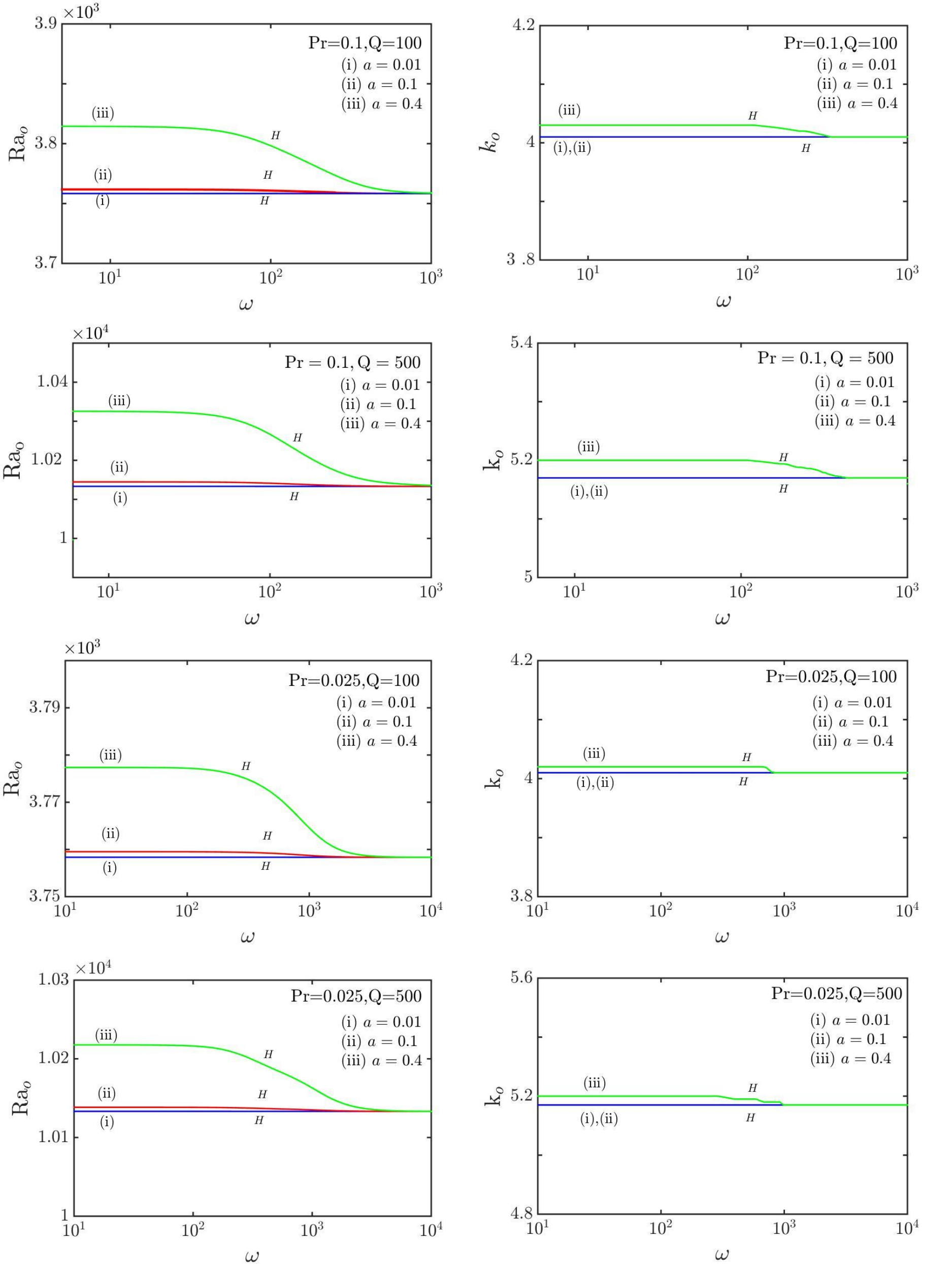}
	\caption{(Colour online) Variations of critical values 
$\mathrm{Ra}_o (\omega)$ (left column) and  $k_o (\omega)$ (right column) with $\omega$ for a given set of $\mathrm{Q}$ and $\mathrm{Pr}$. The magnetoconvection is always harmonic ($\gamma = 0$) for smaller values of the  modulation amplitude $a$. Blue (black), red (grey) and light green (light grey) curves are for (i) $a=0.01$, $0.1$ and $0.4$, respectively.}
	\label{low_a&w}
\end{figure}
				
Variations of $\mathrm{Ra}_o$ and $k_o$ with the frequency $\omega$ are displayed in Fig.~\ref{low_a&w} for smaller values of $a$  and a  given set of values of $\mathrm{Q}$ and $\mathrm{Pr}$. Curves showing the variation of $\mathrm{Ra}_o$ with $\omega$ are displayed in the left column, while those showing the variation of $k_o$  with $\omega$ are displayed in the second column. Curves in the first two rows are for $\mathrm{Pr} = 0.1$. The dimensionless modulation frequency ($\omega$) is varied from $5$ to $10^3$. This corresponds to the variation of actual frequency from $0.1$ Hz to $24$ Hz for a fluid of thickness $2$ mm. For a fluid of thickness of $1$ cm, it corresponds to a variation of actual frequency from $0.005$ Hz to $0.96$ Hz. The onset of magnetoconvection is delayed at lower values of modulation amplitude and frequency. However, the behaviour here is different to one observed for smaller value of $\mathrm{Q}$ and for very small values of $\omega$ (Fig~\ref{low_a_Q50}). The threshold $\mathrm{Ra}_o (\mathrm{Q}, \mathrm{Pr}, a, \omega)$ increases as $a$ becomes larger in the range of dimensional frequencies considered here. The wavy flow at the onset always oscillates synchronously for smaller values of $a$ and $\omega$ for $\mathrm{Q}=100$. As $a$ is raised to higher values, the onset of magnetoconvection is further delayed. For higher values of $\mathrm{Q}$, the threshold $\mathrm{Ra}_o$ becomes higher. The threshold for magnetoconvection  $\mathrm{Ra}_o (\mathrm{Q}, \mathrm{Pr})$ remain almost constant for a reasonable large window of dimensionless frequencies. It finally approaches to its value for the stationary magnetoconvection $\mathrm{Ra}_c (\mathrm{Q})$, as $\omega \rightarrow \infty$. The critical wave-number also displays a similar behaviour. The value of $k_o$ is larger than its value for stationary convection $k_c$ and its value changes very little in the frequency window. As $\omega$ is increased significantly, its value starts decreasing and finally approaches to $k_c$, the critical wave-number for the stationary magnetoconvection. Curves shown in the last two rows of Figure~\ref{low_a&w} are for  mercury ($\mathrm{Pr} =0.025$, $\nu = 1.14 \times 10^{-7} m^2/s$). In this case $\omega$ has been varied from $10$ to $10^4$. For a layer of mercury of thickness $2$ mm, it corresponds to a variation of $\tilde{f}$ approximately from $0.05$ Hz to $45$ Hz. Similarly, for $1$ cm thick layer of mercury, this corresponds to a variation of $\tilde{f}$ approximately from $2 \times 10^{-3}$ Hz to $2$ Hz. The behaviour observed in mercury, at smaller values of modulation amplitude and frequency, is qualitatively similar.  

\begin{figure}[!]
	\includegraphics[height=!, width=8.0cm]{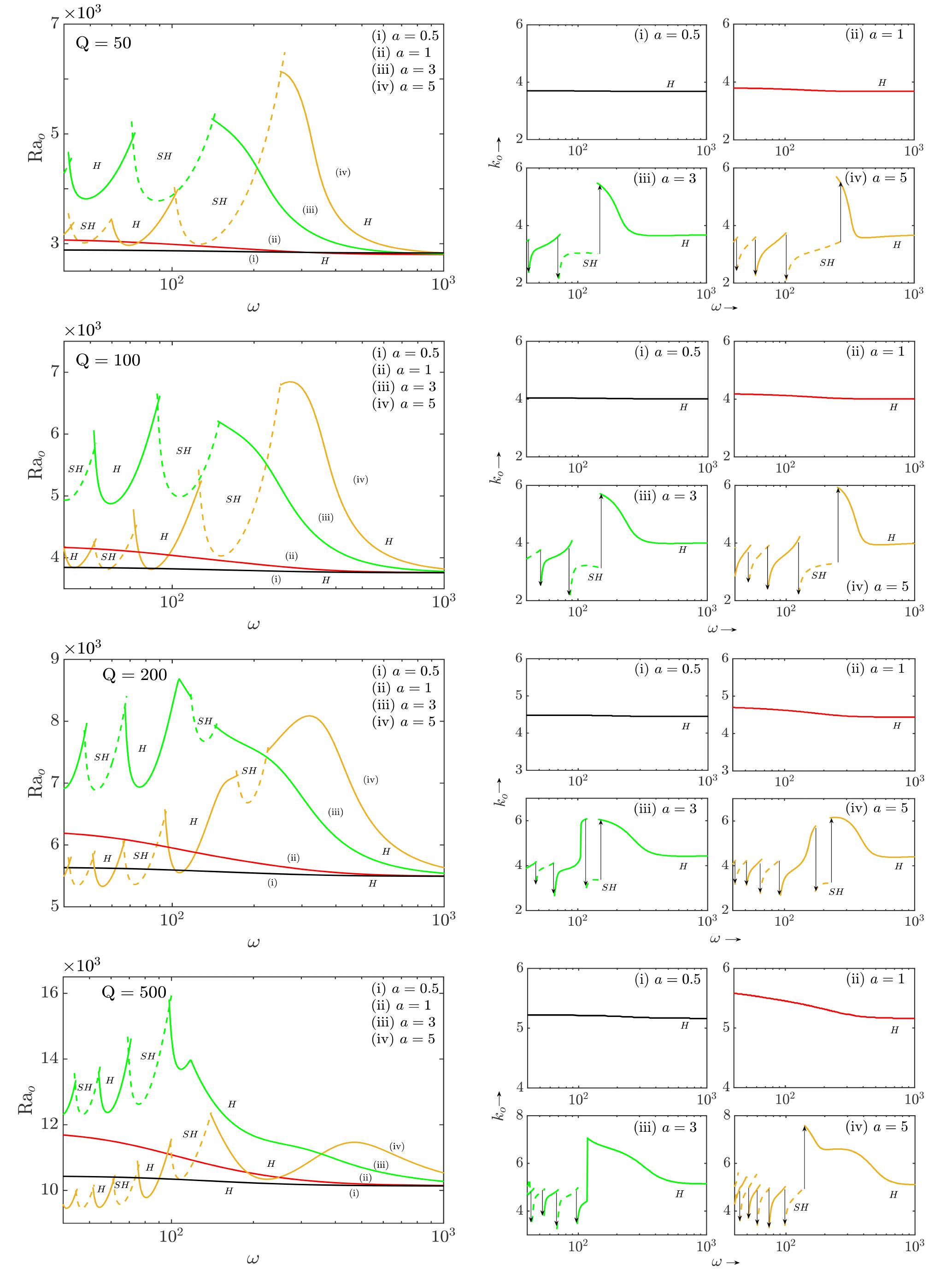}
	\caption{(Colour online) Variations of $\mathrm{Ra}_o$ (left column) and   $k_o$ (right column) with $\omega$ are plotted for $\mathrm{Pr} =0.1$ and for several values of $a$.  Viewgraphs in different rows are for different values of $\mathrm{Q}$. Solid curves are for the harmonic flow ($\gamma = 0$) and dashed curves are for the subharmonic flow ($\gamma=1/2$). Curves of the same colour (or grey level) correspond to the same value of amplitude of modulation, $a$.}
	\label{Pr01_wR_wk}
\end{figure}

Figure~\ref{Pr01_wR_wk} displays the variations of $\mathrm{Ra}_o$ and $k_o$ with $\omega$ for for $\mathrm{Pr} = 0.1$ for different sets of values for  $a$ and $\mathrm{Q}$. The plots are shown for non-dimensional frequencies between $40$ and $10^3$. As discussed above, for a fluid of $\mathrm{Pr} = 0.1$ and thickness $2.0$ mm $\le d \le 1.0$ cm, the actual frequency of temperature modulation would be in a range from $0.038$ Hz to $1$ Hz. Curves displayed in the top and the bottom rows of Fig.~\ref{Pr01_wR_wk} are for $\mathrm{Q} = 50$ and $500$, respectively. Similarly, the curves shown in the second and third rows from the top are for $\mathrm{Q}=100$ and $200$, respectively. The variation of $\mathrm{Ra}_o$ with $\omega$ is plotted in the left column, and that of $k_o$ with $\omega$ is shown in the right column. The dashed curves show the threshold for subharmonic (SH) oscillation of the magneto-convective flow, while the solid curves show the threshold for harmonic (H) oscillation of the magneto-convective flow. Curves of the same colour (or the same grey level) are for the same value of $a$. Each of the viewgraphs is for a different value of $\mathrm{Q}$. Each of them display four curves for different values of $a$. The magnetoconvection shows harmonic oscillations at the onset for $a\le 1.0$ for all values of  $\mathrm{Q}$ considered here. The threshold $\mathrm{Ra}_o^H$ for $\mathrm{Q}=50$ and $a=0.5$ is $2885$ at $\omega = 40$ and its value approaches to the threshold for stationary magnetoconvection $\mathrm{Ra}_c (\mathrm{Q}=50)=2802$, as $\omega \rightarrow \infty$. As $a$ is raised from $0.5$ to $1.0$, keeping all other parameters fixed, the magnetoconvection remains harmonic but $\mathrm{Ra}_o^H$ becomes larger at smaller values of $\omega$ [see the red (dark grey) curves in the left column]. So the small amplitude temperature modulation stabilises the conduction state at lower frequencies for $a \le 1.0$. For relatively larger values of the modulation amplitude, the nature of magneto-convective flow depends on the modulation frequency as well as the value of $\mathrm{Q}$. 
	
The magneto-convective flow oscillates subharmonically for $a=3.0$ and $\mathrm{Q} =50$, and the threshold $\mathrm{Ra}_o^{SH}$ is much higher at $\omega = 40$ compared to threshold at smaller values of $a$ [see the green (grey) curve]. The first bi-critical point, where $\mathrm{Ra}_o^H = \mathrm{Ra}_o^{SH} = 4490$, appears at $\omega= 42$. Two sets of waves with different wave-numbers ($k_o^{H}$ and $k_o^{SH}$) are excited simultaneously for $a=3.0$ and $\mathrm{Q} = 50$ at this frequency.  With slight increase in $\omega$, $\mathrm{Ra}_o^{H}$ becomes lower than $\mathrm{Ra}_o^{SH}$. This makes the excitation of harmonic  magnetoconvection preferable at the onset. The value of $\mathrm{Ra}_o^H$ first decreases, reaches a minimum and then increases, as $\omega$ is raised. The second bi-critical point occurs at $\omega = 72$ in this case. As the frequency is raised slightly above this value, the threshold for excitation of subharmonic oscillation $\mathrm{Ra}_o^{SH}$ becomes lower than $\mathrm{Ra}_o^H$. The new threshold $\mathrm{Ra}_o^{SH}$ also decreases first, reaches a minimum, and then increases, as $\omega$ is raised. The third bi-critical point appears at $\omega=142$.  Further increase in $\omega$ makes the minimum of another harmonic instability zone lower than  $\mathrm{Ra}_o^{SH}$. At higher values of $\omega$, the magneto-convective flow is synchronous with the modulation at the instability onset. For $a=5.0$, the qualitative behaviour is quite similar. However, the flow is harmonic at the onset for $\omega =40$. This values of threshold $\mathrm{Ra}_o$ is  smaller for $a=5.0$ than those for $a=3.0$ for $\omega < 200$. At higher frequencies ($\omega \gg 200$), the threshold for all cases approach their values for stationary magnetoconvection.  At any given frequency ($\omega > 200$), the threshold is higher for larger values of $a$, if $\mathrm{Pr}$ and $\mathrm{Q}$ are kept fixed.
			
The variation of $k_o$ with the dimensionless frequency $\omega$ is displayed in the right column for different values of $\mathrm{Q}$.  All curves in the top right view-graph are for $\mathrm{Q}=50$ and for different values of $a$. For $a=0.5$, the variation of critical wave-number for harmonic waves, $k_o^{H}$, is insignificant (the black curve) as is the case for for the variation of $\mathrm{Ra}_o^H$ (top left view-graph). For $a=1.0$, the variation of $k_o^{H}$ is visible at lower frequencies [see the red (dark grey) curve].  However, for $a=3.0$, both harmonic and subharmonic oscillations of magneto-convective flow are possible for different frequency windows. The fluid motion is subharmonic at $\omega =40$ and the critical wave-number $k_o^{SH}$ increases initially. When the threshold for excitation of harmonic waves becomes lower than that for subharmonic waves, the wave-number corresponding to the minimum of marginal curve for harmonic instability is selected. This leads to a jump in the wave-number to a lower value [see the green (grey) curve]. The selected wave-number $k_o^{H}$ now slowly increases with $\omega$. Another bicritical point becomes possible when $\mathrm{Ra}_o^{SH}$ becomes equal to $\mathrm{Ra}_o^{H}$. With even slight increase in $\omega$ further,  subharmonically oscillating fluid flow is again expected to be appear at the instability onset.  At a relatively higher value of $\omega$, the fluid flow oscillates harmonically with the external modulation. The wave-number always shows a jump at a bicritical point.  The arrows are drawn at the locations of bicritical points and their directions indicate an increase or a decrease of the wave-number $k_o$.  For $a=5.0$, the variation of $k_o$ with $\omega$ shows the similar behaviour but the number of bi-critical points and their locations on the frequency axis are different.

Viewgraphs in the second, third and fourth rows from the top in  Fig.~\ref{Pr01_wR_wk} are for $\mathrm{Q}=100$,  $200$ and $500$, respectively. There exists low frequency windows at relatively higher values $\mathrm{Q}$, where the magneto-conductive state can either be stabilised or destabilised by modulating the temperature of the lower plate about a mean value. There is an interesting observation at low modulation frequency for $\mathrm{Q} = 500$. For $\mathrm{Q} = 500$ (see the bottom row of Fig.~\ref{Pr01_wR_wk}), both harmonically and subharmonically waves may be excited at lower frequencies for Rayleigh numbers, which are considerably lower than the threshold $\mathrm{Ra}_c (\mathrm{Q})$ for the stationary magnetoconvection in the absence of temperature modulation. The subharmonic manteoconvection may be excited at $\mathrm{Ra}_o^{SH} = 9568$ and harmonic magnetoconvection may be excited at $\mathrm{Ra}_o^H = 9781$. Both these values are much smaller than the threshold for stationary convection ($\mathrm{Ra}_c = 10133$) in absence of temperature modulation. The accurate prediction of possible  stabilisation or destabilisation of the conduction state of modulated Rayleigh-B\'{e}nard magnetoconvection is tricky at lower values of the modulation frequency. At higher frequencies, only the harmonic flow is excited. The threshold $\mathrm{Ra}_o^H$ is higher for larger modulation amplitude. As $\omega \rightarrow \infty$, both $\mathrm{Ra}_o^H$ and $k_o^{H}$ approach the appropriate values for stationary magnetoconduction.

\begin{figure}[!]
 \includegraphics[height=!, width=8.0cm]{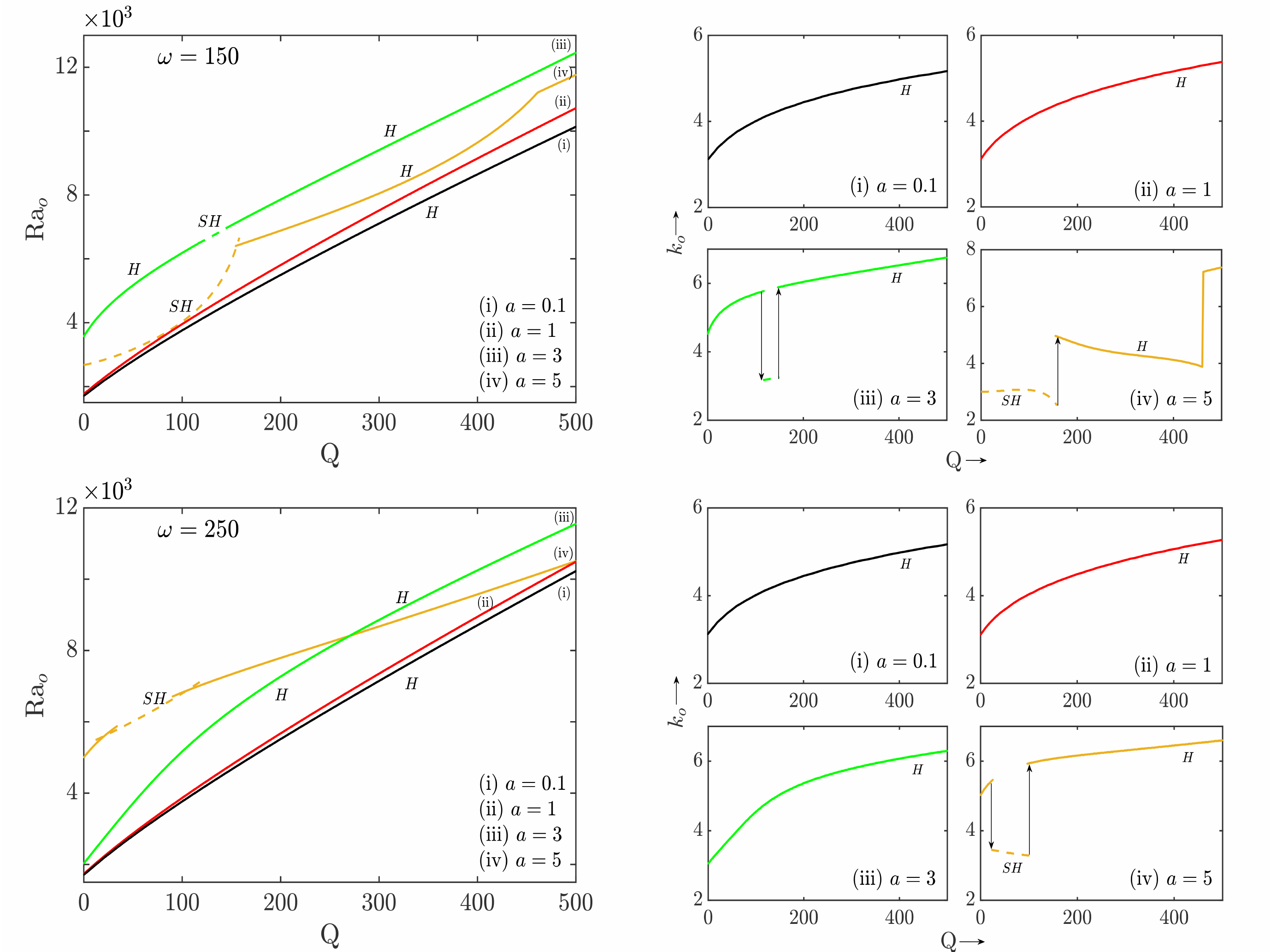}
  \caption{(Colour online) Variations of $\mathrm{Ra}_o$ and $k_o$ with 
	$\mathrm{Q}$ for $\mathrm{Pr}=0.1$. The  upper row displays the 
	 computed curves for $\omega = 150$, while the lower row show the curves for 
	$\omega = 250$. The black, red (dark grey), green (grey) and golden
	(light grey) coloured curves correspond to modulation amplitude $a = 0.1$,
	 $1.0$, $3.0$ and $5.0$, respectively. The solid curves are for harmonic (H)
	 response and dashed curves are for subharmonic (SH) response.}
	 \label{Pr01_QR_Qk}
\end{figure}
		
Effects of the variation of Chandrasekhar's number $\mathrm{Q}$ on the threshold for magneto-convection and the critical wave-number are displayed in Figure~\ref{Pr01_QR_Qk} for $\mathrm{Pr} = 0.1$. The variation of $\mathrm{Ra}_o$ with $\mathrm{Q}$ is shown in the left column, while that of $k_o$ with $\mathrm{Q}$ is displayed in the right column. Curves in the upper row are for modulation frequency $\omega = 150$ and those in the lower row are for $\omega = 250$ for four different values of $a$. For $\omega=150$ and lower values of modulation amplitude ($a \le 1.0$), the harmonic magnetoconvection is excited at the instability onset.  The threshold $\mathrm{Ra}_o^H (\mathrm{Q}$) increases monotonically with $\mathrm{Q}$, as shown by black and red (dark grey) curves (the upper left viewgraph). The variation of $\mathrm{Ra}_o$ with $\mathrm{Q}$ shows interesting behaviour for larger values of $a$. For $a = 3.0$ and $\omega = 150$, the magnetoconvection appears as harmonic waves for lower values of $\mathrm{Q}$. Curves showing the variation of $\mathrm{Ra}_o$ with $\mathrm{Q}$ show either concavity or convexity. Solid green (grey) portion of the curve, which shows the variation of $\mathrm{Ra}_o^H$ with $\mathrm{Q}$, is convex. As $\mathrm{Q}$ is raised above a critical value (here, $\mathrm{Q}=118$), the magnetoconvection becomes subharmonic at the primary instability. Dashed green (grey) portion of the curve, which shows the variation of $\mathrm{Ra}_o^{SH}$ with $\mathrm{Q}$, becomes concave. Again at $\mathrm{Q} = 147$, the magneto-convective flow at the instability onset is harmonic. For $\mathrm{Q} > 148$, $\mathrm{Ra}_o^H$ increases with $\mathrm{Q}$ as almost linearly till $\mathrm{Q} = 500$. For $\mathrm{Pr} =0.1$, $a=3.0$ and $\omega =150$, there are two bi-critical points. The first one at $\mathrm{Q} = 118$, where $\mathrm{Ra}_o^H$ $=$ $\mathrm{Ra}_o^{SH}$ $=$ $6509$.  The second bi-critical point is observed at $\mathrm{Q} = 147$, where $\mathrm{Ra}_o^H$ $=$ $\mathrm{Ra}_o^{SH}$ $=$ $7002$. 

On the other hand, the magnetoconvection appears as subharmonically oscillating flow for $a=5.0$, $\omega=150$ and $\mathrm{Q} < 155.6$ at the onset. There appears a bi-critical point for $\mathrm{Q} = 155.6$, where $\mathrm{Ra}_o^{SH}$ $=$ $\mathrm{Ra}_o^H$ $=$ $6417$. For $155.6 < \mathrm{Q} < 500$, the harmonically oscillating flow is observed at the instability onset. Variations of both $\mathrm{Ra}_o^{SH}$ [shown by the dashed golden (light grey) curve] and $\mathrm{Ra}_o^{H}$ [shown by the solid golden (light grey) curve] with $\mathrm{Q}$ are concave in this case. There is another bi-critical point at $\mathrm{Q}=461$, where two synchronous waves have the same threshold ($\mathrm{Ra}_o^{H1}=\mathrm{Ra}_o^{H2}=1.121\times10^4$). Besides, the value of threshold for $a=5.0$ is lower than that for $a=3.0$ with all other parameters held fixed. This is possible as the minimum of different instability zones become the lowest as different control parameters are varied. They move upward or downward in the $\mathrm{Ra}$-$k$ plane differently. It is hard to predict which one would be the lowest at relatively smaller modulation frequencies without performing Floquet analysis.  Four curves in the right column of the first row show the variation of $k_o$ with $\mathrm{Q}$. For smaller values of the modulation amplitude ($a \le 1.0$), $k_o^H (\mathrm{Q})$ increases monotonically with $\mathrm{Q}$. However, the curve showing variation of $k_o$ with $\mathrm{Q}$ shows two jumps at two bi-critical points for $a=3.0$ and two jumps for $a=5.0$, one of which shows a transition from subharmonically oscillating flow to harmonically oscillating flow and the other shows a transition from one set of harmonic flow to another set of harmonic flow. 

For $\mathrm{Pr} = 0.1$, $\omega = 250$ and $\mathrm{Q} \le 500$, the magnetoconvection always appears as harmonically oscillating flow for $a < 3.0$ (the viewgraph at bottom left). However for $a=5.0$, the fluid flow at primary instability may be subharmonic in a window of $\mathrm{Q}$ ($20 < \mathrm{Q} < 104$). The $\mathrm{Ra}_o^H$-$\mathrm{Q}$ curve is convex, while the $\mathrm{Ra}_o^{SH}$-$\mathrm{Q}$ curve is concave. For higher value of $\mathrm{Q}$, the excited wavy flow oscillates harmonically. However the threshold for $a=5.0$ is lower than that for $a=3.0$. Four curves in the lower right viewgraphs describe the variation of $k_o$ with $\mathrm{Q}$. There are no bi-critical points for $a \le 3.0$ but two bi-critical points appear for $a=5.0$.
\begin{figure}[H]
 \includegraphics[height=!, width=8.0cm]{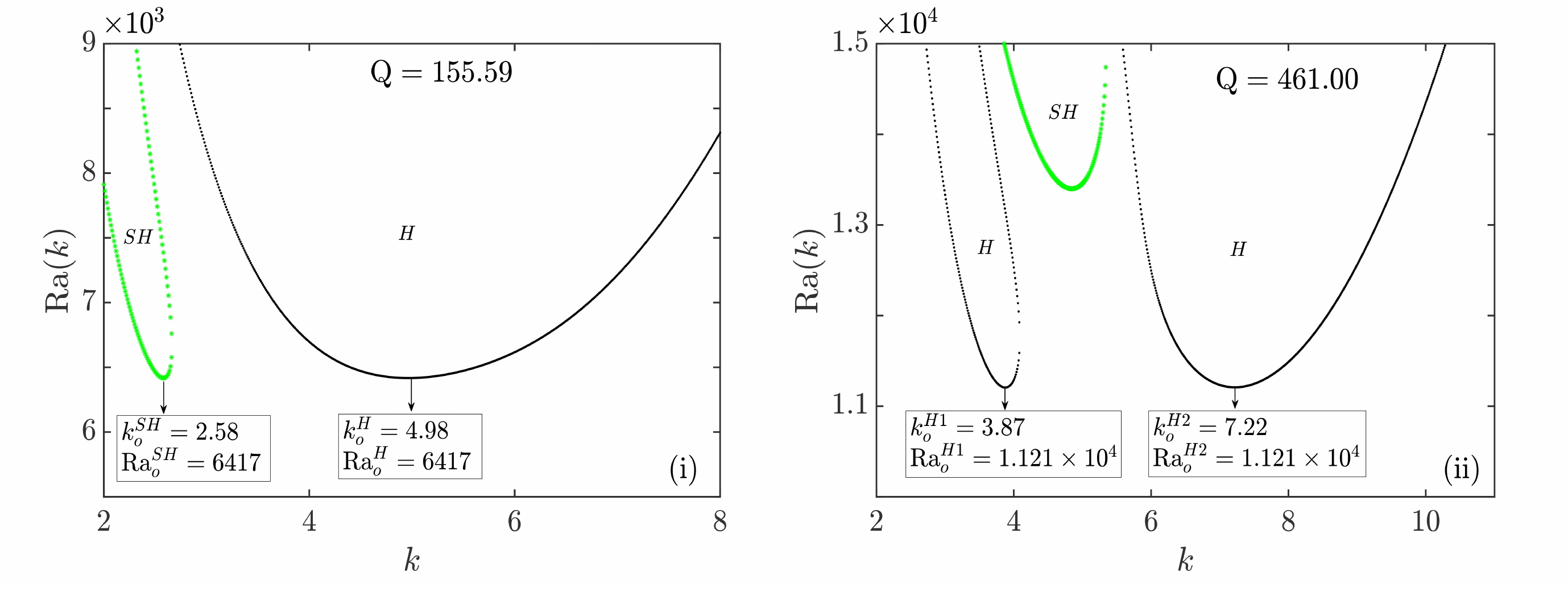}
  \caption{(Colour online) Two different types of bicritical points for 
  $\mathrm{Pr}=0.1$, $a=5.0$ and $\omega=150$. The left viewgraph displays a bicritical point showing the simultaneous excitation of subharmonic and  harmonic fluid flows at the instability onset for $\mathrm{Q}=155.59$. The right viewgraph shows a bicritical point for $\mathrm{Q}=461$, where two sets of harmonically oscillating waves of different wave-numbers are simultaneously possible. The regions inside black curves are instability zones for harmonic magnetoconvection, while the region inside green (light grey) curve is instability zone for subharmonic magnetoconvection.}
	 \label{Pr01_Rk_Q}
\end{figure}

Figure~\ref{Pr01_Rk_Q} displays two types of bi-critical points for $\mathrm{Pr}=0.1$, $a=5.0$ and $\omega=150$. For $\mathrm{Q}=155.59$, a bi-critical point is observed at $\mathrm{Ra}_o=6417$. This involves one set of subharmonic waves of wave-number $k_o^{SH}=2.58$ and another set of harmonic waves of wave-number $k_o^{H}=4.98$. As $\mathrm{Q}$ is raised to relatively higher value, both subharmonic and harmonic instability zones move to higher values of $\mathrm{Ra}$ and $k$ and a new harmonic zone appears for smaller values of $k$. As $\mathrm{Q}$ is raised further, all these tongues move  upward in the $\mathrm{Ra}$-$k$ plane. However, the tongue-like region for subharmonic instability moves upward faster than tongue-like regions for harmonic instability. As a consequence, there is possibility of two different harmonic instability zones with their minima at the same value of $\mathrm{Ra}$. The right viewgraph of Fig.~\ref{Pr01_Rk_Q} shows such a situation for $\mathrm{Q}=461.0$. Two sets of harmonically oscillating waves with wave-numbers $k_o^{H1}=3.87$ and $k_o^{H2}=7.22$ are simultaneously excited at 
$\mathrm{Ra}_o=1.121\times10^4$. This kind of bi-critical point is new and not reported earlier.

\begin{figure}[H]
 \includegraphics[height=!, width=8.0cm]{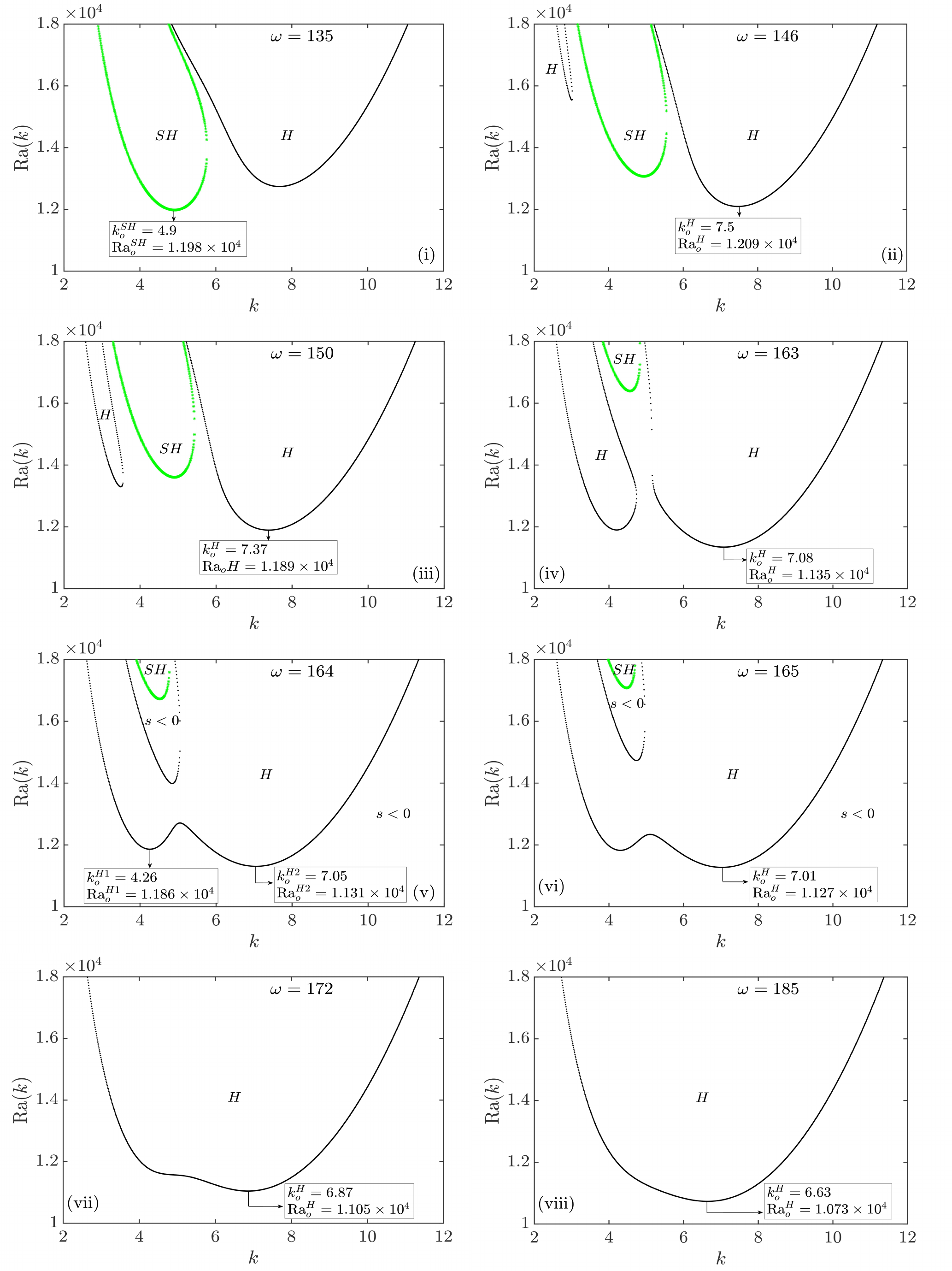}
  \caption{(Colour online) Instability zones for $\mathrm{Pr} = 0.1$, 
   $\mathrm{Q} = 500$ and $a = 5$ with different values of $\omega$. The regions
    inside black curves are instability zones for harmonic magnetoconvection and
    the regions inside green (light grey) curves are instability zones for 
    subharmonic magnetoconvection.}
	 \label{Pr01_Rk_w}
\end{figure}

Effects of the modulation frequency $\omega$ on the instability zones for $\mathrm{Pr} =0.1$ and $\mathrm{Q}=500$ are shown in Fig.~\ref{Pr01_Rk_w}. The first two lowest tongues are shown for $\omega=135$ (see, the top left viewgraph). The magnetoconvection appears as subharmonically oscillating waves with a wave-number $k_o^{SH} = 5.01$ for the modulation frequency $\omega=135$. At slightly higher value of the modulation frequency, i.e., at $\omega =146$ (the top right viewgraph), the instability zone for harmonically oscillating waves become the lowest. The magnetoconvection at the instability onset is then synchronous with the external modulation. The critical wave-number $k_o^{H}=7.5$ becomes higher than its value, $k_o^{SH} = 4.9$, at $\omega=135$. The significant change in the wavelength of the waves would be immediately noticeable at the transition point. In addition, a new harmonic instability zone appears at lower values of $k$, which was not existing at $\omega=135$. The instability region for subharmonically oscillating flow is surrounded from two sides by instability zones for harmonically oscillating flows in the $\mathrm{Ra}$-$k$ plane. An interesting phenomenon occurs, as $\omega$ is raised further. Both the regions for harmonic instability move downward in the $\mathrm{Ra}$-$k$ plane. The newly created instability zone moves faster than the older zone for harmonic instability. The region for subharmonic instability moves upward, as $\omega$ is raised. As a result, the first two lowest marginal curves correspond to excitation of harmonic  oscillation of the magneto-convective flow.
		
We observed interesting behaviour, as $\omega$ is raised further. The  subharmonic instability zone, which is sandwiched between two harmonic instability zones in the $\mathrm{Ra}-k$ plane, moves further up, while  the harmonic instability zones widens (see the viewgraph for $\omega = 163$). A slight increase in $\omega$ makes two marginal curves for harmonic instability merge to form a single marginal curve with two local minima. The merger of two harmonic zones at $\omega = 164$ is displayed Fig.~\ref{Pr01_Rk_w} (see the the left viewgraph in the third row). The region between the upper boundary of the marginal curve for harmonic instability and the lower boundary of the marginal curve for subharmonic instability corresponds to stable conduction state, where the growth rate of all perturbative fields is negative ($s < 0$). Raising $\omega$ to higher values pushes the subharmonic zone further up. This makes the possibility of observing subharmonically oscillating fluid flow at the onset of magnetoconvection impossible. The marginal curve for harmonic instability attains one global minimum for $\omega \ge 185$. The temperature modulation leads to the possibility of only harmonically oscillating magneto-convective waves at higher values of $\mathrm{Q}$ even at moderate value of $\omega$. 
		
\begin{figure}[H]
 \includegraphics[height=!, width=8.0cm]{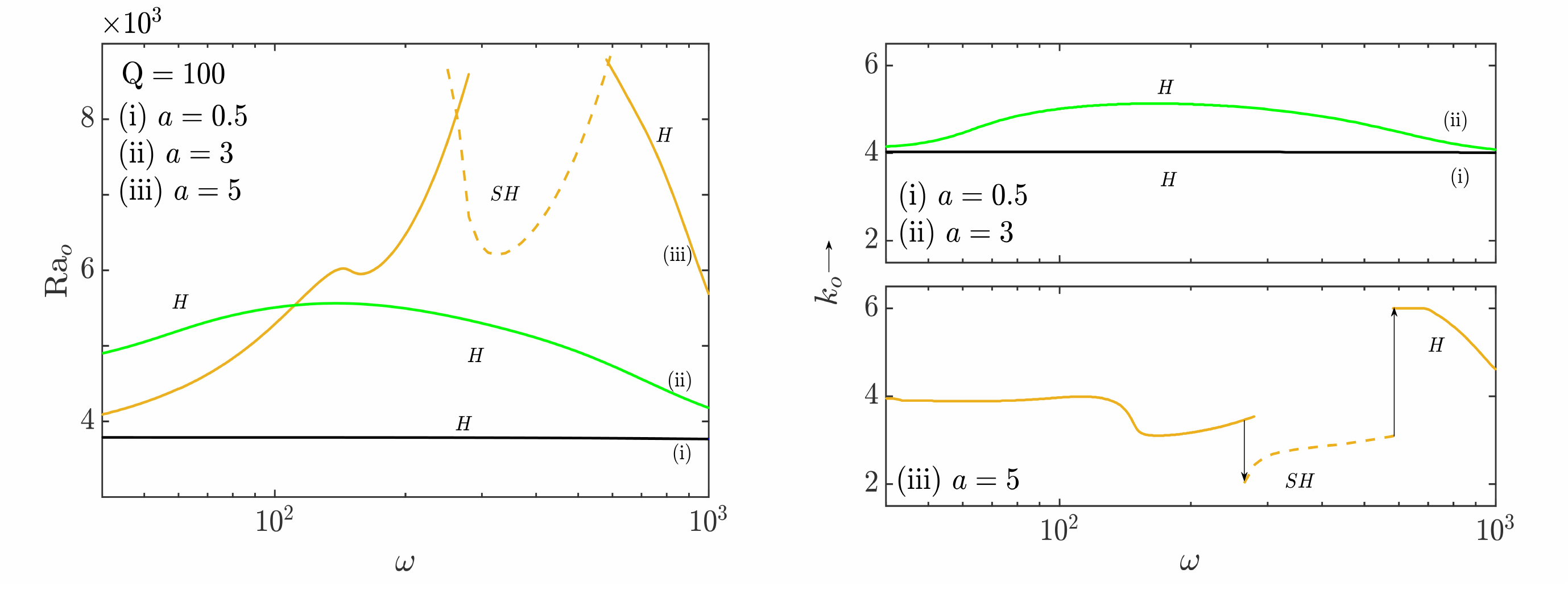}
  \caption{(Colour online) Variations of $\mathrm{Ra}_o (\omega)$ and $k_o (\omega)$ with the modulation frequency are shown for liquid mercury ($\mathrm{Pr}=0.025$) for $\mathrm{Q}=100$. The left viewgraph shows the variation of $\mathrm{Ra}_o$ with $\omega$, while the right viewgraph displays the variation of $k_o$ with $\omega$. The black, green (grey) and golden yellow (light grey) curves correspond to $a=0.5$ $a=3.0$ and $a=5.0$, respectively. Solid and dashed curves stand for harmonic and subharmonic oscillations, respectively.}
	 \label{Pr025_wR_wk}
\end{figure}

\begin{figure}[!]
 \includegraphics[height=!, width=8.0cm]{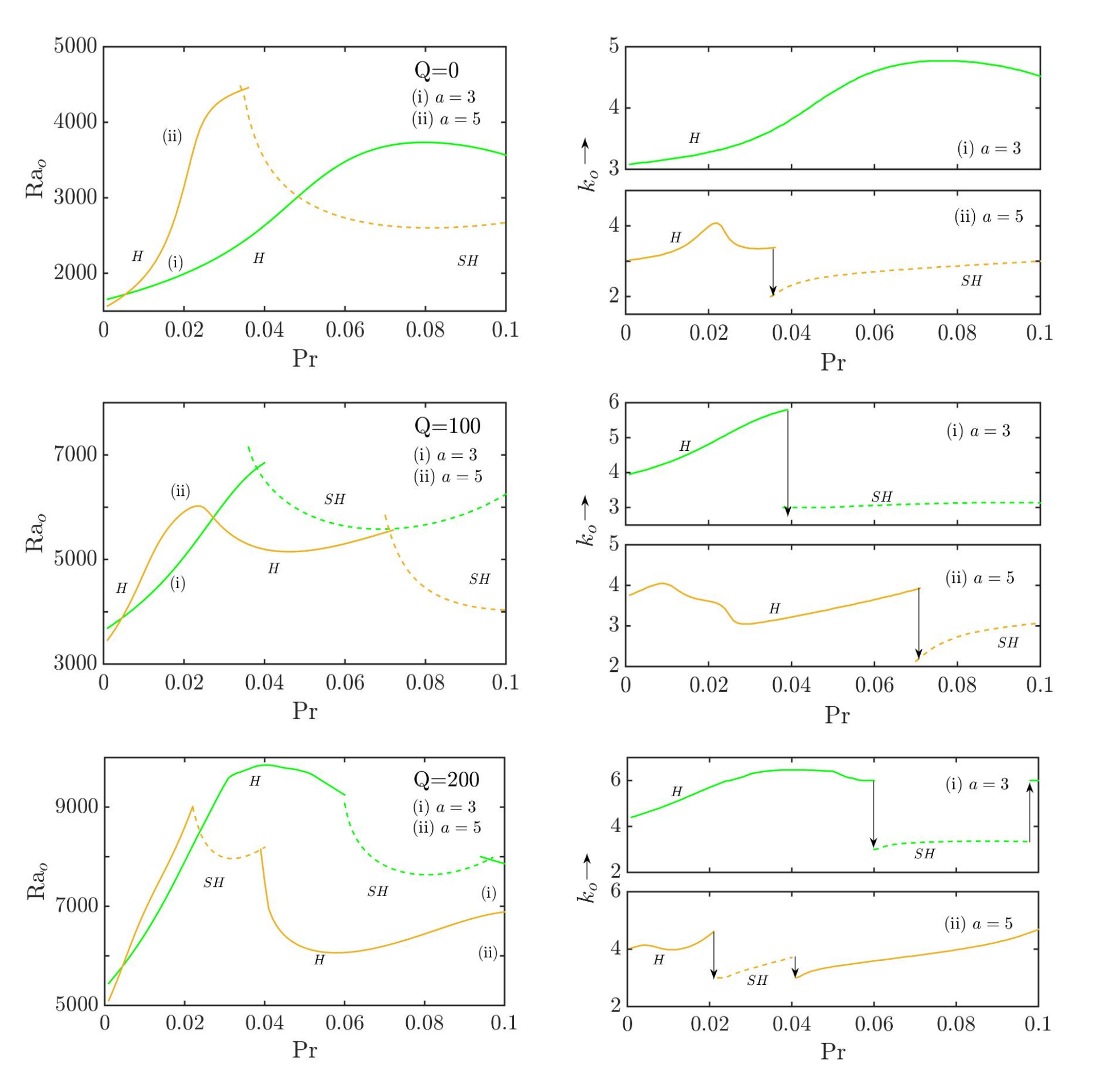}
  \caption{(Colour online) Effects of variation of the Prandtl number $\mathrm{Pr}$ on  critical values $\mathrm{Ra}_o (\mathrm{Pr})$ and $k_o(\mathrm{Pr})$ are displayed.  Viewgraphs in different rows are for different values of $\mathrm{Q}$. The curves are for $\omega = 150$ and two different values of $a$: (i) $a=3.0$ [curves in green (grey) colour] and (ii) $a=5.0$ [curves in golden yellow (light grey) colour]. The solid and dashed curves are for harmonic and subharmonic oscillations, respectively.}
	\label{w150_pR_Pk}
\end{figure}
		
Figure~\ref{Pr025_wR_wk} displays the variations of critical values of $\mathrm{Ra}_o$ and $k_o$ (the right viewgraph) with modulation frequency $\omega$ in liquid mercury ($\mathrm{Pr} = 0.025$) for $\mathrm{Q}=100$. The black, green (grey) and golden (light grey) coloured curves are for the modulation amplitude $a=0.5$, $a=3.0$ and $5.0$, respectively. Fluid flow at the instability onset always oscillates harmonically for $a\le 3.0$. However the critical value of $\mathrm{Ra}_o$ shows a maximum at $\omega = 138$. There is almost no variation in the critical values  of $\mathrm{Ra}_o$ and $k_o$ with $\omega$ for much lower values of modulation amplitude (e.g., $a=0.5$). The green (grey) curve, which shows the variation of $\mathrm{Ra}_o^H$ with $\omega$, is slightly convex. This is so as the instability zone for harmonically oscillating waves in the $\mathrm{Ra}-k$  first moves upwards and towards higher values of $k$ and then starts coming downward slowly and leftwards, as $\omega$ is raised continuously. The variation of critical wave-number $k_o^H$ confirms this behaviour. Black and green (grey) curves in the upper viewgraph on the right column show the variation of $k_o$ with the modulation frequency $\omega$ for $a=0.5$ and $3.0$, respectively.

For larger values of $a$, the behaviour is completely different (see the curve for $a=5.0$ in the left viewgraph). The excited flow at the onset of magnetoconvection oscillate  harmonically for lower values of $\omega$ ($\le 260$), as shown by the solid golden (light grey) curve. The oscillation of fluid flow becomes subharmonic for moderate values of value of $\omega$ ($260 \le \omega \le 588$), as shown by the dashed golden (light grey) curve. Again for $\omega > 588$, the magnetoconvection excited at the onset always leads to harmonically oscillating flow, if other parameters are kept fixed.  At bi-critical points observed at $\omega = 260$ and $588$, the fluid flow shows harmonic as well as subharmonic oscillations at the primary instability. The variation of the wave-number with $\omega$ is displayed for this case ($a=5.0$) in the lower viewgraph in the right column. The curve shows jumps in the wave-number at the bi-critical points.  The variation of $\mathrm{Ra}_o$ and $k_o$ depend on $\mathrm{Pr}$, if all other parameters are held at fixed values. 
		
The role of Prandtl number $\mathrm{Pr}$ on the threshold $\mathrm{Ra}_o$ and the critical wave number $k_o$ is also investigated. Figure~\ref{w150_pR_Pk} displays the variations of $\mathrm{Ra}_o$ and $k_o$ with $\mathrm{Pr}$ for modulation frequency $\omega =150$.  The variation of $\mathrm{Ra}_o$ with $\mathrm{Pr}$ is plotted in the left column for different values of $\mathrm{Q}$. The same for $k_o$ is plotted in the right column. A range of Prandtl number ($0 < \mathrm{Pr} \le 0.1$) is considered.  Only the harmonically oscillating flow is possible at the instability onset in absence of any external magnetic field ($\mathrm{Q} = 0$) and modulation amplitude $a=3.0$, as shown by green (grey) solid curve in the upper left viewgraph, for the whole range of $\mathrm{Pr}$ considered. The threshold $\mathrm{Ra}_o^H$ and the corresponding wave number $k_o^H$ both increase monotonically with $\mathrm{Pr}$, as shown by the green (grey) solid curves in the upper row.  However, both the harmonically and sub-harmonically oscillating waves are likely to be excited for $a=5.0$. The harmonic oscillation of fluid flow is observed at the onset of magnetoconvection for $\mathrm{Pr} < 0.034$, while the sub-harmonic oscillation of the flow is expected at the onset for $\mathrm{Pr} > 0.034$. A jump in the selected wave number is observed, if $\mathrm{Pr}$ is raised above  $0.034$.  There is a bi-critical point at $\mathrm{Pr} = 0.0346$, where $\mathrm{Ra}_o^H=\mathrm{Ra}_o^{SH}= 4430$. For $0.05 < \mathrm{Pr} < 0.1$, the threshold for excitation of waves with $a=3.0$ is larger that for $a=5.0$. Curves in the middle row are for $\mathrm{Q} = 100$ and $\omega = 150$. New bicritical points appear at $\mathrm{Pr}=0.038$ for $a=3.0$ and at $\mathrm{Pr}=0.071$ for $a=5.0$. The corresponding jumps in $k_o$ at a bi-critical point is shown in the viewgraphs at middle right. For a range of $\mathrm{Pr}$ ($0.005<\mathrm{Pr}< 0.027 $) the threshold to excite magnetoconvection is lower for $a=3.0$. For $\mathrm{Pr} < 0.005$ and $\mathrm{Pr} > 0.027$, however, the threshold for excitation of waves is lower for $a=5.0$.  As $\mathrm{Q}$ is raised further, the more number of bi-critical points becomes possible. Curves in the lower row are for $\mathrm{Q} = 200$ and $\omega=150$. There are two bi-critical points for $a=3.0$ and $a=5.0$ in this case. The first one occurs for $a=3.0$ at $\mathrm{Pr}=0.06$, where $\mathrm{Ra}_o^H=\mathrm{Ra}_o^{SH}= 9247$. The second one occurs for $0.096$, where $\mathrm{Ra}_o^H$ $=$ $\mathrm{Ra}_o^{SH}$ $=$ $7939$. The first one occurs for $a=5.0$ at $\mathrm{Pr}=0.022$, where $\mathrm{Ra}_o^H=\mathrm{Ra}_o^{SH}= 9000$. The second one occurs for $0.039$, where $\mathrm{Ra}_o^H$ $=$ $\mathrm{Ra}_o^{SH}$ $=$ $8142$. The variation of $\mathrm{Ra}_o$ with $\mathrm{Pr}$ is non-monotonous. Similarly, the increase of modulation amplitude $a$ with other parameters maintained at fixed values may either increase or decrease the threshold for generation of wavy flow. The minimum of different marginal curves in the $\mathrm{Ra}$-$k$ plane becomes the global minimum, as a parameter is varied continuously. The movement of different tongue-shaped instability zone is also different. This makes the variation of critical Rayleigh number with a parameter non-monotonic.

\section{Conclusions}
Floquet analysis of Rayleigh-B\'{e}nard magnetoconvection under a time-periodic temperature modulation shows interesting results. As the Rayleigh number becomes larger than a threshold value $\mathrm{Ra}_o (\mathrm{Q}, \omega, \mathrm{Pr})$, the oscillatory magnetoconvection  is excited. The magneto-convective flow may oscillate either subharmonically or harmonically with the frequency of modulation, $\omega$, for smaller values of $\omega$. Several bi-critical points involving two sets of wavy flows of different wave-numbers are possible for smaller values of $\omega$: one set oscillates subharmonically and the other set  oscillates harmonically.  The magneto-convective flow is found to be always synchronous with the modulation for larger value of $\omega$.  The threshold for generation of magneto-convective waves $\mathrm{Ra}_o$ varies non-monotonically with lower frequencies of modulation but at finite values of the amplitude of modulation, $a$. A new type of bi-critical point is possible for larger value $\mathrm{Q}$ when two sets of harmonic oscillations with two different wave lengths  may be excited simultaneously at the onset of magnetoconvection.  For larger values of Chandrasekhar number $\mathrm{Q}$ and moderate values of $\omega$, two marginal instability zones for harmonic instability merge to form a single marginal curve with two local minima. They may have interesting consequences on nonlinear behaviour. \\   

\noindent
{\bf Acknowledgement:}	
We acknowledge the partial support from the Grant No. EMR/2016/000185 of SERB Project (India). Discussions with S. P. Khastgir (IIT Kharagpur) were beneficial.

\end{document}